\documentclass[ reprint,
superscriptaddress,
 amsmath,amssymb,
 aps,
prd,
floatfix,
]{revtex4-2}

\usepackage{graphicx}% Include figure files
\usepackage{dcolumn}% Align table columns on decimal point
\usepackage{bm}% bold math
\usepackage{xcolor} % Required for coloring text
\usepackage{subcaption} % For subfigures
\newcommand{\gev}{\textrm{GeV}}
\newcommand{\mev}{\textrm{MeV}}

\newcommand{\nue}{\nu_{\textrm{e}}}
\newcommand{\nuebar}{\bar{\nu_{\textrm{e}}}}
\newcommand{\numu}{\nu_{\mu}}
\newcommand{\numubar}{\bar{\nu_{\mu}}}
\newcommand{\p}{\textrm{p}}

\newcommand{\pip}{\pi^+}

\newcommand{\piz}{\pi^0}

\newcommand{\deltapp}{\Delta^{++}}

\newcommand{\ccopiop}{\textrm{CC}1\pi1\textrm{p}}

\newcommand{\numuccopi}{\nu_{\mu}\textrm{CC}1\pi}

\newcommand{\numuccopiop}{\nu_{\mu}\textrm{CC}1\pi1\textrm{p}}

\newcommand{\dat}{\delta\alpha_\textrm{T}}

\newcommand{\dpt}{\delta p_\textrm{T}}

\newcommand{\dptt}{\delta p_\textrm{TT}}

\newcommand{\vecpsum}{\vec{p}_{\textrm{sum}}}

\newcommand{\vecppicom}{\vec{p}_{\pi}^{(\textrm{COM})}}
\newcommand{\thetacom}{\theta_{\textrm{COM}}}
\newcommand{\ecom}{E_{\textrm{COM}}}
\newcommand{\thetaadt}{\theta_{\textrm{Adt}}}
\newcommand{\thetapidel}{\theta_{\pi\Delta}}
\newcommand{\thetaR}{\theta_{\pi\textrm{R}}}

\newcommand{\vecpdel}{\vec{p}_{\Delta}}

\newcommand{\plong}{p_{\textrm{long}}}

\newcommand{\enu}{E_{\nu}}
\newcommand{\vecpp}{\vec{p}_\textrm{p}}
\newcommand{\vecppi}{\vec{p}_{\pi}}
\newcommand{\vecppp}{\vec{p}_\textrm{p}^{\prime}}
\newcommand{\vecppip}{\vec{p}_{\pi}^{\prime}}
\newcommand{\vecppirest}{\vec{p}_{\pi}^{(0)}}

\newcommand{\MA}{M_{\textrm{A}}}
\newcommand{\dcp}{\delta_\textrm{CP}}

\newcommand{\genie}{\texttt{GENIE}}
\newcommand{\geoa}{\texttt{G18-0a}}
\newcommand{\getwoa}{\texttt{G18-02a}}
\newcommand{\geta}{\texttt{G18-10a}}
\newcommand{\getb}{\texttt{G18-10b}}
\newcommand{\gz}{\texttt{G24-0}}
\newcommand{\gc}{\texttt{G24-c}}

\newcommand{\chindf}{\chi^2/{\textrm{NDF}}}

\newcommand{\scfigwid}{0.45}
\newcommand{\dbfigwid}{0.45}
\newcommand{\qufigwid}{0.23}

\begin{document}

\preprint{APS/123-QED}

\title{\textbf{Centre-of-momentum Variables in $\nu_\mu$CC1p1$\pi$} 
}% 

\author{Weijun Li}
 \affiliation{University of Oxford, Dept. of Physics, Oxford OX1 3RH, UK}%Lines break automatically or can be forced with \\
  \email{weijun.li@physics.ox.ac.uk}

\date{\today}% It is always \today, today,
             %  but any date may be explicitly specified

\begin{abstract}
This study introduces a novel set of variables, namely the centre-of-momentum variables, $\thetacom$ and $\ecom$, designed to isolate final-state interactions (FSI) from other aspects of neutrino-nucleus interactions. 
Through detailed simulation studies, this work demonstrates the ability of these variables to distinguish FSI contributions with minimal dependence on the nuclear initial state and, practically, on the neutrino flux, highlighting their potential for advancing FSI modeling. 
With high-purity neutrino-hydrogen interaction selections, $\thetacom$ offers the first opportunity for a direct cross-comparison among different neutrino cross-section experiments.
\end{abstract}

%\keywords{Suggested keywords}%Use showkeys class option if keyword
                              %display desired
\maketitle

%\tableofcontents
\section{Introduction}
Significant efforts have been devoted to measuring Charge-Parity (CP) violation in the neutrino sector through long-baseline (LBL) experiments. 
In these experiments, CP violation is quantified by the difference between neutrino oscillation and anti-neutrino oscillation. 
LBL experiments quantify $\dcp$ by comparing the energy spectra of $\nue$ and $\nuebar$ oscillated from $\numu$ and $\numubar$, respectively.
There are two major challenges to this measurement. 
Firstly, the far detector of an LBL experiment is by design hundreds of kilometres away from the neutrino source, which unavoidably leads to low statistics.
Secondly, as the energy of each incoming neutrino is unknown and thus the type of the individual neutrino bound-nucleon interaction is also unknown, oscillation predictions have to take the form of energy spectra for a chosen final state topology.

Large-scale experiments, such as Hyper-Kamiokande~\cite{Hyper-Kamiokande:2018ofw} and the Deep Underground Neutrino Experiment (DUNE)~\cite{DUNE:2015lol,DUNE:2016evb,DUNE:2016hlj,DUNE:2016rla,DUNE:2021tad}, address the first challenge by constructing gigantic far detectors to increase event rates.
To match the reduction of statistical uncertainties, it is critical to develop advanced neutrino-nucleus interaction models or to better constrain existing ones to minimize the systematic uncertainties arising from the second challenge.
The neutrino-nucleus interaction is a convolution of multiple processes: the nucleon initial state (IS), the neutrino-nucleon interaction, and final state interactions (FSI). 
In particular, due to FSI, an event topology, e.g. CC$0\pi$, does not correspond only to a type of neutrino-nucleon interaction, e.g. quasi-elastic interactions, but also contains contributions from other interactions such as resonance production, because FSI could produce or absorb additional pions.
Thus, accurate $\dcp$ measurements depend heavily on neutrino interaction models estimating contributions of different neutrino-nucleon interactions to final-state topologies~\cite{NuSTEC:2017hzk}.

To better understand the complex neutrino-nucleus interactions, new or upgraded experiments with sophisticated detectors have commenced to explore a larger interaction kinematic phase space and to collect a significantly larger amount of data. 
For instance, the Tokai-to-Kamioka (T2K) experiment~\cite{T2K:2011qtm} has upgraded its near detector (ND) and started data collection in June 2024. 
The Super Fine-Grain Detector (SFGD), part of the T2K ND upgrade~\cite{T2K:2019bbb}, provides improved proton detection with lower thresholds, higher resolution, and greater efficiency. 
Meanwhile, the Short-baseline Near Detector (SBND)~\cite{MicroBooNE:2015bmn} has also begun operations in 2024. 
It is a new LArTPC with an active mass of 112 ton placed at $110$ m from the neutrino source. 
Due to its large active mass and proximity to the source,  it is expected to collect a huge number of neutrino interaction events each year.  
On one hand, the expanded kinematic phase space allows for measuring new variables. 
On the other hand, the influx of high-quality data offers an ideal testing ground for novel measurement techniques.
Taking advantage of these advancements, it is timely to explore new ideas involving pions in the final states, given the broad energy spectrum of the DUNE beamline, which includes substantial contributions from resonance production comparable to quasi-elastic interactions, 

One effective method of utilizing the near detector data is to constrain model parameters through tuning.
Successful examples~\cite{GENIE:2021zuu,GENIE:2021wox,GENIE:2022qrc} have shown improved data-Monte Carlo (MC) agreement after tuning existing models using various combinations of measurements from different experiments. 
% However, the neutrino-nucleus interaction is a convolution of multiple processes: the nucleon initial state (IS), the neutrino-nucleon interaction, and FSI. 
One difficulty of tuning is that many variables are affected by all processes of the complicated neutrino-nucleus interaction, making it both challenging to study the different models in isolation and numerically expensive and cumbersome to tune all processes at once.
Nuclear effects, such as IS and FSI, occur within the nucleus and remain unobservable with current detectors, making them a major source of systematic uncertainties.
Cleverly constructed variables, such as Transverse Kinematic Imbalance (TKI)~\cite{Lu:2015hea, Lu:2015tcr} or Generalized Kinematic Imbalance (GKI)~\cite{MicroBooNE:2023krv}, are sensitive to nuclear effects, and past measurements have successfully constrained models~\cite{GENIE:2024ufm}. 
While TKI is sensitive to both IS and FSI, except $\dat$, which is predominantly sensitive to FSI but is affected by small uncertainties in the neutrino direction, new variables like $\plong$ \cite{Baudis:2023tma} are designed to be sensitive to specific nuclear effects, such as the removal energy. 

Having more specialized measurements, like $\plong$, can further fine-tune our models, especially in light of the improved detection capabilities. 
This work proposes a new set of variables, called center-of-momentum (COM) variables, for charge current single pion single proton ($\ccopiop$) events, timely for the increasingly precise measurements with pions in the final states
COM variables enable more focused studies of FSI by differentiating between FSI models independently of IS models.

This paper will elaborate on the concept of the COM variables and present MC analysis results focusing on the COM angle and demonstrating its ability to distinguish FSI models and its independence from IS.

\section{The COM Variables}
\label{sec:com}
When a neutrino has sufficient energy, it can excite a nucleon into a resonance state, for example the $\deltapp$. 
This process is referred to as a resonance (RES) interaction.
Since $\deltapp(1232)$, simply referred to as $\deltapp$ unless stated otherwise, is one of the most commonly observed resonances in neutrino experiments, this work focuses on it as an example.
Nevertheless, the concept presented here is equally applicable to other resonances, and the methodology can be easily generalized.

\begin{figure}[ht!]
    \centering
    \includegraphics[width=\linewidth]{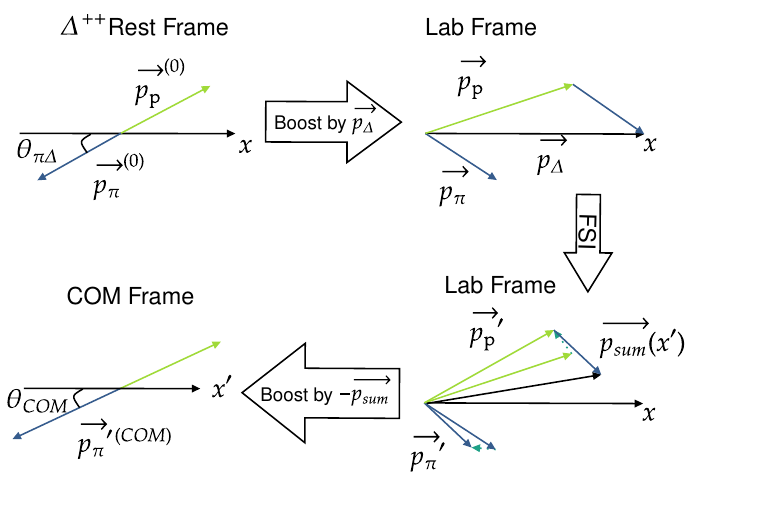}
    \caption{Schematic illustration of the COM angle. Without FSI, $\vecpsum=\vecpdel$ and the lab frame and the $\deltapp$ rest frame can be transformed into each other by $\vecpdel$. With FSI, $\vecpsum \neq \vecpdel$, the $\deltapp$ rest frame is not accessible, but the lab frame can be boosted into the COM frame using $\vecpsum$. Hence, the major difference of the COM frame from the $\deltapp$ rest frame is caused by FSI. }
    \label{fig:COM-diagram}
\end{figure}

The resonance decays rapidly, before leaving the nucleus, via the process
\begin{equation}
	\deltapp \rightarrow \pip + \p.
\end{equation}
In the $\deltapp$ rest frame, as illustrated on the top left in Fig.~\ref{fig:COM-diagram}, the kinematics of this two-body decay are well-defined and the proton and pion are emitted back-to-back.
The pion decay angle, $\thetapidel$, is defined as the angle between $\vecppirest$ and the $x$-axis, which is taken to align with $\vecpdel$, the momentum of $\deltapp$ in the lab frame. 
$\thetapidel$ is a resonance property that follows an underlying distribution defined by various models~\cite{Rein:1987cb,Kabirnezhad:2017jmf,Kabirnezhad:2020wtp,Kabirnezhad:2022znc}.

The kinematics in the lab frame are related to those in the $\deltapp$ rest frame by a boost with $\vecpdel$, as depicted on the top right of Fig.~\ref{fig:COM-diagram}. 
Without FSI, $\vecpdel$ equals the sum of $\vecpp$ and $\vecppi$.
However, the target materials of modern day detectors are mainly comprised of nuclei with multiple nucleons and FSI alters the kinematics of the hadrons considerably, as illustrated by the dotted arrows in the bottom right of Fig.~\ref{fig:COM-diagram}.

The altered momenta, $\vecppp$ and $\vecppip$, are the ones measured by detectors. 
Their sum generally differs from $\vecpdel$.
Thus, the $\deltapp$ rest frame with its simple kinematic relations becomes inaccessible.
Nevertheless, the system can be boosted to the proton-pion COM frame using $\vecpsum$, as depicted in the bottom left of Fig.~\ref{fig:COM-diagram}, where the $x^{\prime}$-axis is taken to align with the $\vecpsum$ direction in the lab frame.
Similarly, a pion decay angle, $\thetacom$, can be defined between, $\vecppicom$, the pion momentum in the COM frame, and the $x^{\prime}$-axis.
Due to FSI, $\thetacom$ typically differs from $\thetapidel$. 

The COM frame coincides with the $\deltapp$ rest frame only in the absence of FSI.
Thus, the strength of FSI dictates the deviation of $\thetacom$ from $\thetapidel$. 
In practice, the measured $\thetacom$ serves as a probe for studying FSI. 
$\thetapidel$, being a rest-frame property of $\deltapp$, is independent of the resonance's momentum, neutrino energy, and IS.
In neutrino event generators, $\thetacom$ deviates from $\thetapidel$ only due to FSI, which is independent of neutrino energy and IS.
Therefore, $\thetacom$ retains these important independencies.
As different resonances could have different decay properties, $\thetaR$, the similarly defined pion decay angle of a higher resonance, $R$, generally differs from $\thetapidel$. 
Hence, $\thetacom$, a superposition of $\thetapidel$ and all possible $\thetaR$, will deviate from $\thetapidel$, when more resonances become energetically possible with an increase in neutrino energy. 
This deviation will be further elaborated in the Sec.~\ref{sec:dis}.

Moreover, the total energy in the COM frame, $\ecom$, will be equal to the mass of the resonance in the absence of FSI.
Cutting on events with total energy far from the rest mass peak of the resonance will select events with minimal FSI effects, such as the $\nu$-H events. 
Also, an upper cut on $\ecom$, e.g. $\ecom<1330~\gev$, can help to select events with a $\deltapp$ resonance, as the $\deltapp$ rest mass is $1232~\mev$.
However, this work will focus on the COM angle only, and the investigation of the versatile utilitiy of $\ecom$ will be conducted in future works.

Up to this point, the discussion of resonance decay has been entirely general and can thus be applied to and validated by various experiments, including hadron scattering measurements. 
However, there are features specific to neutrino experiments. 
First, the resonance is generated through neutrino-nucleon interactions, which involve both vector and axial currents. 
Second, the interaction occurs within the nuclear medium. 
This work focuses on the latter, with the application of COM variables to the former reserved for future research.

While the COM angle may appear conceptually similar to the reconstructed Adler angle, $\thetaadt$, in neutrino experiments~\cite{Sanchez:2015yvw}, there are significant differences. 
Notably, the COM frame is reconstructed exclusively from hadronic kinematics, whereas the reconstructed Adler frame relies on leptonic kinematics and assumes stationary nucleons. 
Consequently, the reconstruction of the Adler frame is implicitly influenced by IS effects, whereas the COM frame's hadronic variables are impacted by final state interactions (FSI). 
Therefore, $\thetaadt$, being a hadronic variable in the Adler frame, is influenced by both IS and FSI. 
Additionally, the necessity of reconstructing the neutrino energy renders $\thetaadt$ sensitive to neutrino flux uncertainties, which are among the largest systematic uncertainties in neutrino measurements~\cite{T2K:2019yqu,T2K:2021naz,MicroBooNECollaboration:2024gvg,NOvA:2023uxq,MINERvA:2022djk}.

\section{Analysis result}
\label{sec:ana}
In the absence of existing measurements of the COM angle, MC studies were conducted to evaluate its potential advantages—namely, robustness against IS effects and sensitivity to FSI effects.
MC samples were generated using \genie~ \cite{Andreopoulos:2009rq, GENIE:2021npt} for the T2K beam and target, unless otherwise noted. 
Several \genie~ tunes are employed in this section, and their model details are summarized in Table~\ref{tab:genie-tunes}.
Each sample consists of $600,000$ muon neutrino events. 
For the T2K flux on carbon using \gz, about $70\%$ of the events are charge‐current (CC) events, of which on average $16\%$ are CC single pion ($\numuccopi$) events and $9\%$ are CC single pion single proton ($\numuccopiop$) events.
A selection of $\numuccopiop$ events was used to generate the plots.
Based on preliminary MC studies of the T2K upgraded near detector, approximately $1.8\times10^{21}$ POT will yield this number of events, corresponding to less than one year of data taking in the $\numu$ mode with the J-PARC beam upgrade~\cite{T2K:2019eao}.
A few plots were generated using the MINERvA low-energy (LE) flux, which has a neutrino energy peak around $3.5~\gev$.
The collected MINERvA LE data contain approximately $300,000$ $\numu$ events—half the size of the simulated event sample used in this analysis—and are therefore used only as a proof of concept.
Although the MINERvA flux used in this analysis is the LE flux, similar performance is expected for the medium-energy flux, which has a neutrino energy peak at $6~\gev$ and provides more than 10 times the data.
\begin{table}[h]
    \centering
    \begin{tabular}{c|c|c|c|c|c}
     CMC                  &  IS  &  QE                & 2p2h         & RES & FSI\\
     \colrule
     \geoa    &  RFG         &   LS               & Dytman       & RS  & hA\\
     \getwoa    &  RFG         &   LS               & Dytman       & BS  & hA\\
     \geta    
                          &  LFG         &  Valencia          & Nieves       & BS  & hA\\
     \getb    &  LFG         &  Valencia          & Nieves       & BS  & hN\\
     \gz    &  SF-CFG      &  Valencia          & SuSAv2       & BS  & hA\\
     \gc    &  SF-CFG      &  Valencia          & SuSAv2       & BS  & hA(tuned)\\
    \end{tabular}
    \caption{
        The full tune names are: 1) \geoa: \texttt{G18\_01a\_02\_11b}; 2) \getwoa: \texttt{G18\_02a\_02\_11b}; 3) \geta: \texttt{G18\_10a\_02\_11b}; 4) \getb: \texttt{G18\_10b\_02\_11b}; 5) \gz: \texttt{G24\_20i\_00\_000}; 6) \gc: \texttt{G24\_20i\_06\_22c}.
        The four nuclear IS models are relativistic Fermi gas (RFG), local Fermi gas (LFG), spectral-function-like CFG (SF-CFG)~\cite{sfcfg-talk,sfcfg-GitHubCommit,GENIE:2021npt}. 
    The respective cross-section models are: 1) Quasielastic (QE) - Llewellyn-Smith (LS)~\cite{LlewellynSmith:1971uhs} and Valencia~\cite{Nieves:2004wx}; 2) 2 particle 2 hole (2p2h) - Dytman~\cite{genie:2p2h-dytman}, Nieves~\cite{Nieves:2011pp} and SuSAv2~\cite{Gonzalez-Jimenez:2014eqa}; 3) Resonance(RES) - Rein-Sehgal (RS)~\cite{Rein:1980wg} and Berger-Sehgal (BS)~\cite{Berger:2007rq}. 
    The FSI models used are the built-in \genie~ models~\cite{Andreopoulos:2015wxa}: INTRANUKE hA and hN.
    Note that in \texttt{G24\_20i\_06\_22c}~\cite{GENIE:2024ufm}, the hA model has been tuned using TKI data, resulting in parameters that differ from the default version of hA.
    }
    \label{tab:genie-tunes}
\end{table}

\begin{figure}
    \centering
    \begin{subfigure}[ht!]{\scfigwid\textwidth}
        \centering
        \includegraphics[width=\textwidth]{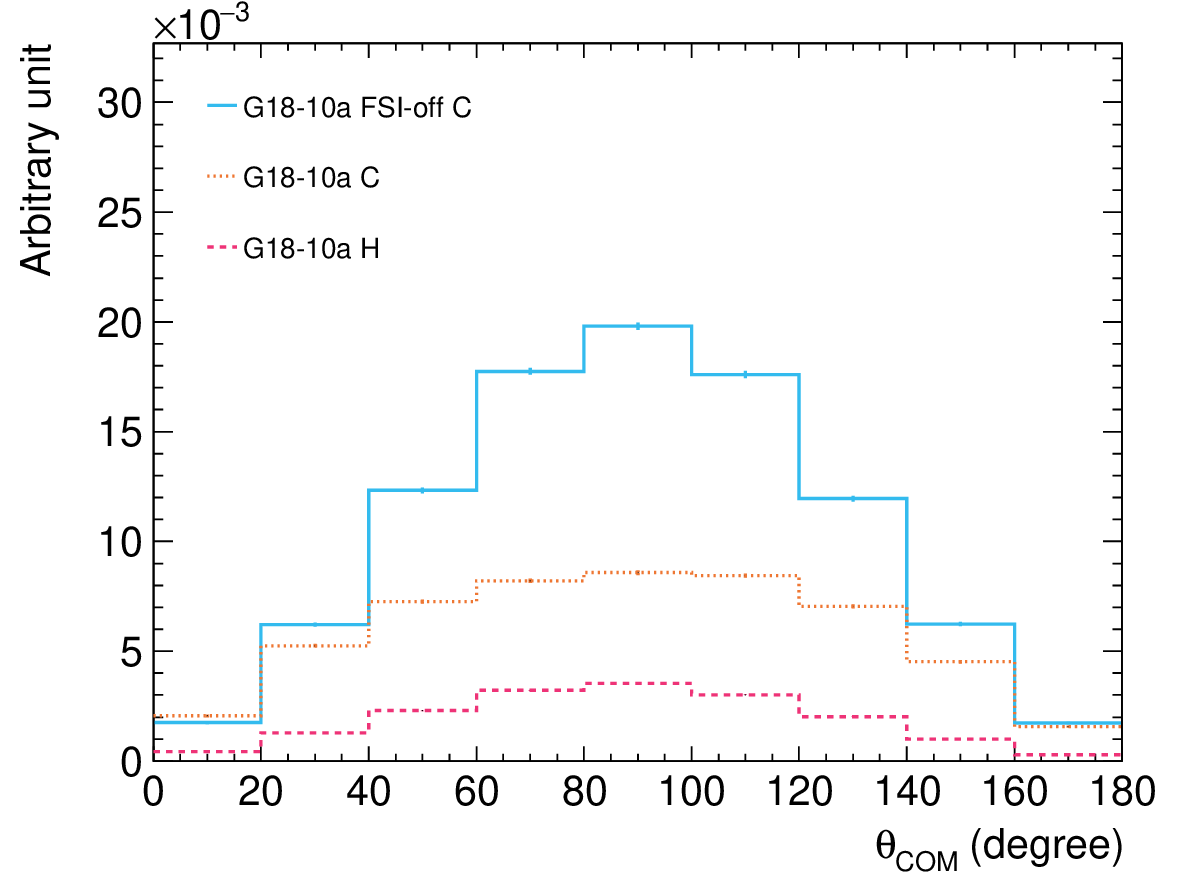} \\
        \caption{Cross-section (per nucleus) normalized $\thetacom$ distribution - $\numuccopiop$ selection.}
        \label{subfig:ch-comp-xnorm}
    \end{subfigure}
    \begin{subfigure}[ht!]{\scfigwid\textwidth}
        \centering
        \includegraphics[width=\textwidth]{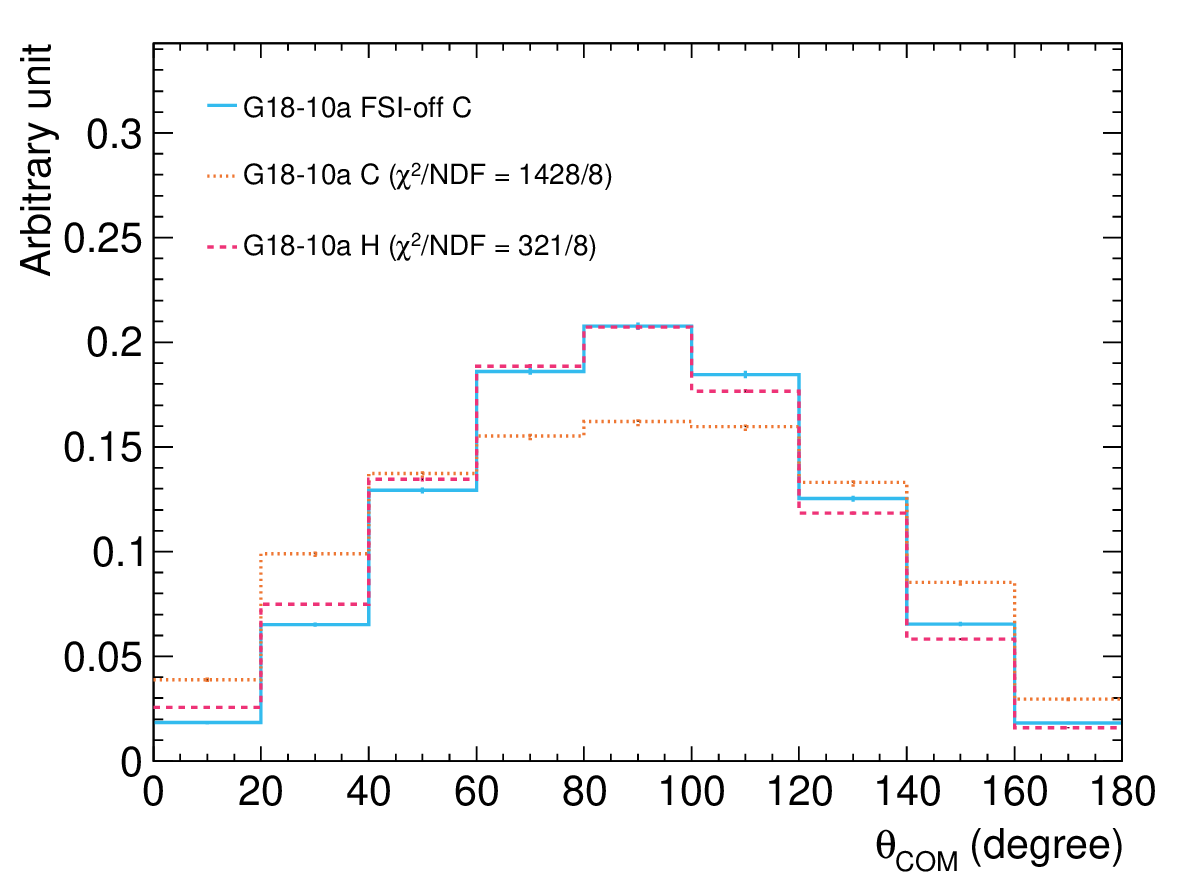}     
        \caption{Area normalized $\thetacom$ distribution - $\numuccopiop$ selection.}
        \label{subfig:ch-comp-com-cc1pi1p}
    \end{subfigure}
    \begin{subfigure}[ht!]{\scfigwid\textwidth}
        \centering
        \includegraphics[width=\textwidth]{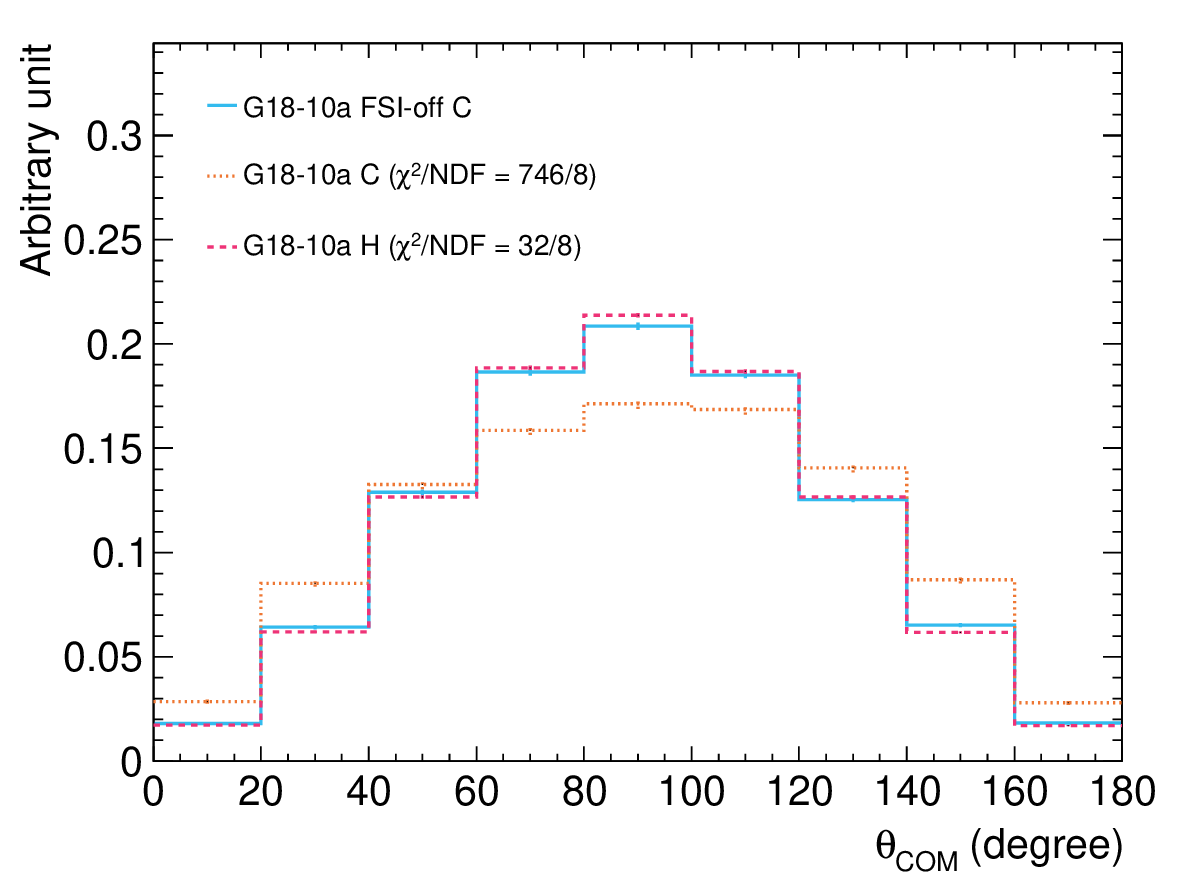}  
        \caption{Area normalized $\thetacom$ distribution - $\numuccopiop$ selection restricted to $\deltapp$ decay only.}
        \label{subfig:ch-comp-com-dpp}
    \end{subfigure}
    \caption{Cross-section normalized (\ref{subfig:ch-comp-xnorm}) and area normalized (\ref{subfig:ch-comp-com-cc1pi1p} and \ref{subfig:ch-comp-com-dpp}) comparisons for FSI on/off for carbon and hydrogen with \geta for $\thetacom$ using the T2K flux. (\ref{subfig:ch-comp-com-dpp}) has a more stringent selection to restrict the resonance to be $\deltapp$.}
    \label{fig:ch-comp}
\end{figure}

The most stringent test of $\thetacom$ with respect to IS effects is to compare the $\thetacom$ distributions for $\nu$-H and $\nu$-C events when FSI is disabled.
The $\nu$-H interaction is free from any nuclear effects, whereas $\nu$-C events without FSI reflect only the impact of IS, including effects such as carbon removal energy and Fermi motion.
FSI effects can be examined by comparing the $\thetacom$ distributions of $\nu$-C events obtained with FSI enabled versus disabled.
When FSI is enabled, two of the major nuclear effects—both IS and FSI—is present.
The cross-section–normalized $\thetacom$ distributions for all three types of events are shown in Fig.~\ref{subfig:ch-comp-xnorm}.

As illustrated in Fig.~\ref{subfig:ch-comp-xnorm}, enabling or disabling FSI has a significant effect on the $\thetacom$ cross-section.
The $\thetacom$ distributions of the two $\nu$-C (FSI-on and FSI-off) samples are considerably larger than that of $\nu$-H, owing to the larger number of nucleons.
Given that the event selection is based on the topology (i.e. $\numuccopiop$), enabling FSI can lead to a considerable decrease in the cross section via pion absorption and via pion‐induced pion production.
Pion charge exchange can also convert $\pip$ to $\piz$, thereby reducing the cross section; however, this type of FSI has a negligible impact for the T2K flux~\cite{GENIE:2024ufm}.
Readers can refer to the literature (e.g. Ref.~\cite{Filali:2024vpy}) for further demonstration of the impact of FSI on the cross section.

To examine the detailed effect of IS on $\thetacom$, the cross-section shape of $\nu$-C events (with FSI enabled and disabled) is compared to that of $\nu$-H in Fig.~\ref{subfig:ch-comp-com-cc1pi1p}.
To better quantify the shape differences, $\chindf$ values (where NDF stands for the number of degrees of freedom) are calculated for all shape‐comparison plots.
Each $\chindf$ value is calculated with respect to the first curve in its respective plot, accounting only for statistical uncertainties.
As shown by the ``\geta~ FSI-off C'' and ``\geta~ C'' curves in Fig.~\ref{subfig:ch-comp-com-cc1pi1p}, enabling FSI leads to a drastic change in the cross-section shape.
In contrast, the $\nu$-H and $\nu$-C (FSI-off) curves are much closer to each other.
This bolsters the claim that FSI has a much stronger impact on the $\thetacom$ distribution than do IS effects.
Nonetheless, the FSI-off $\nu$-C distribution remains considerably different from the $\nu$-H distribution, with a large $\chindf$ value of $321/8$.
If the comparison is restricted to events from $\deltapp$ decay only (based on true information), as shown in Fig.~\ref{subfig:ch-comp-com-dpp}, the difference between the FSI-off $\nu$-C and $\nu$-H $\thetacom$ distributions is significantly reduced, with a $\chindf$ of $32/8$, suggesting that these distributions are almost statistically compatible.
However, the large $\chindf$ observed for $\numuccopiop$ indicates that even without FSI, using carbon instead of hydrogen as the target introduces complexities beyond a simple change in the nucleon kinematic distribution.
It is challenging to pinpoint the exact cause of this large difference, which should be reserved for investigation in future studies.

\begin{figure}
    \centering
    \begin{subfigure}[b]{\dbfigwid\textwidth}
        \centering
        \includegraphics[width=\textwidth]{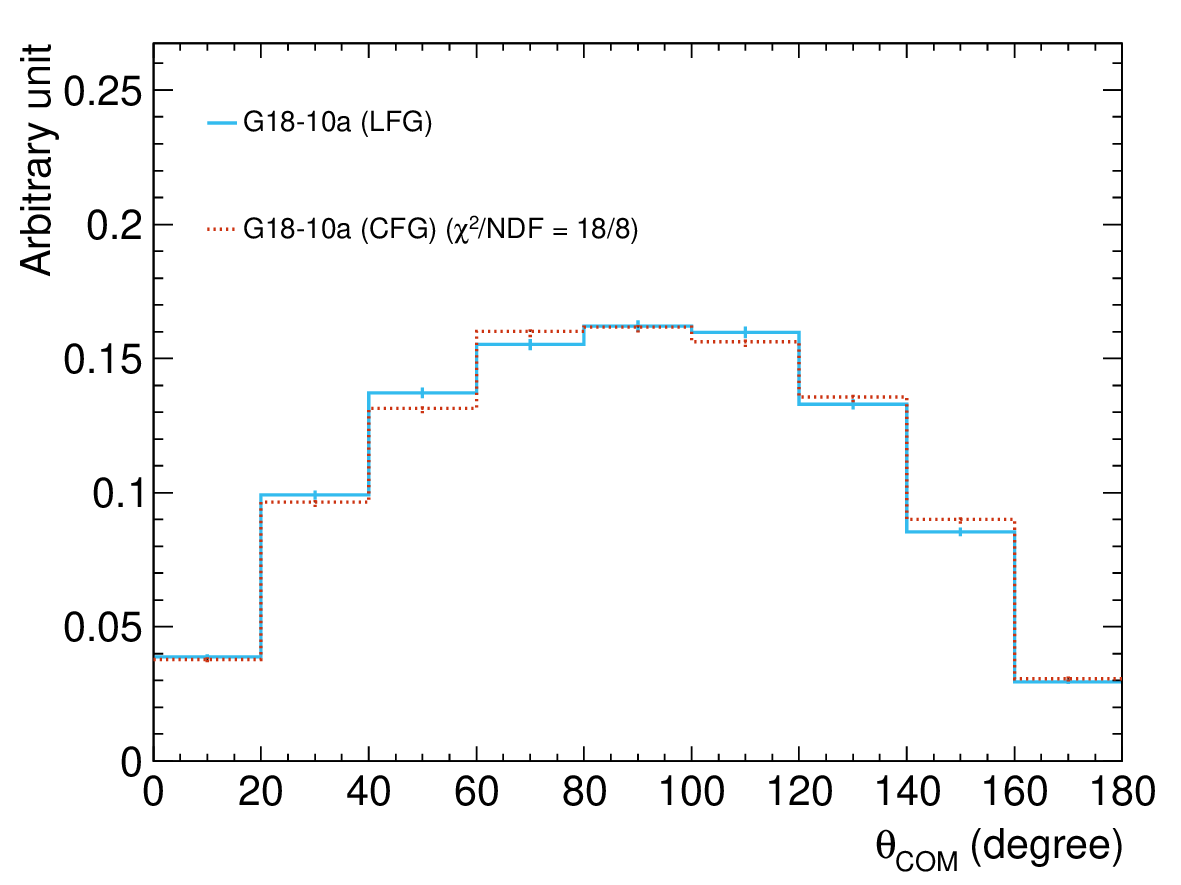}
        \caption{$\thetacom$}
        \label{subfig:10alfg-comp-t2k}
    \end{subfigure}
    \begin{subfigure}[b]{\dbfigwid\textwidth}
        \centering
        \includegraphics[width=\textwidth]{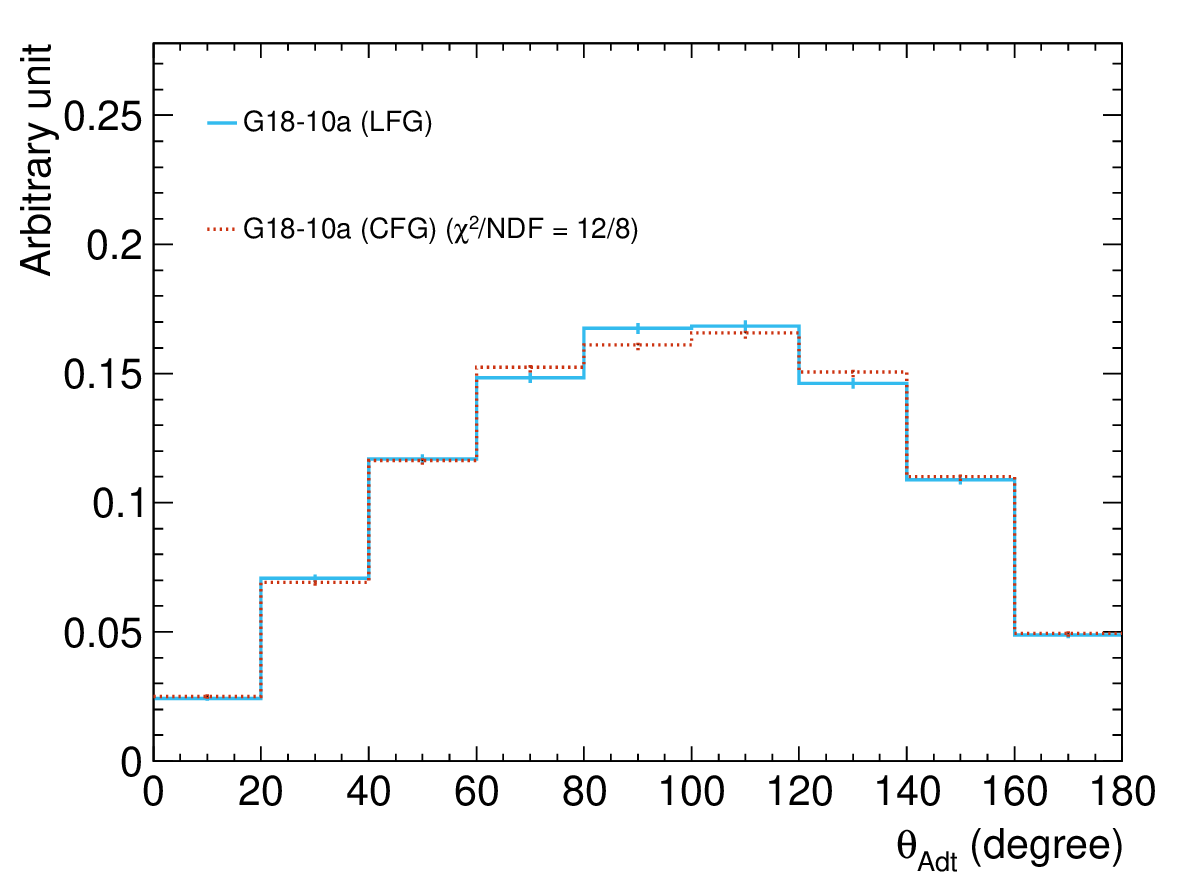}
        \caption{$\thetaadt$}
        \label{subfig:10alfg-comp-t2k-adt}
    \end{subfigure}
    \caption{Area normalized distributions for different IS models, LFG (default) and CFG, with the T2K flux on carbon. The nominal flux is \geta.}
    \label{fig:10a-comp-t2k}
\end{figure}

Because comparing FSI-off carbon with FSI-off hydrogen appears to be too aggressive a test to isolate the impact of IS modeling, a more reasonable approach is to compare different IS models while keeping other factors fixed.
The simulation results for the default \geta~ and its variant using the correlated Fermi gas (CFG) model are shown in Fig.~\ref{subfig:10alfg-comp-t2k}.
The $\chindf$ value of $18/8$ between \geta~ (LFG) and \geta~ (CFG) is indeed small, suggesting that $\thetacom$ is robust against changes in IS models.
Since $\thetaadt$ is conceptually similar to $\thetacom$ but exhibits a larger dependence on IS, it is imperative to compare the $\thetaadt$ distributions for \geta~ (LFG) and \geta~ (CFG) as well, as shown in Fig.~\ref{subfig:10alfg-comp-t2k-adt}, to assess the impact of the IS model change.
Unexpectedly, an even smaller difference between the two curves for $\thetaadt$ is observed in Fig.~\ref{subfig:10alfg-comp-t2k-adt} compared to Fig.~\ref{subfig:10alfg-comp-t2k}, indicating that despite its dependence on IS, changing from LFG to CFG does not distort the shape of $\thetaadt$ appreciably.

\begin{figure}[ht!]
    \centering
    \begin{subfigure}[ht!]{\dbfigwid\textwidth}
        \centering
        \includegraphics[width=\textwidth]{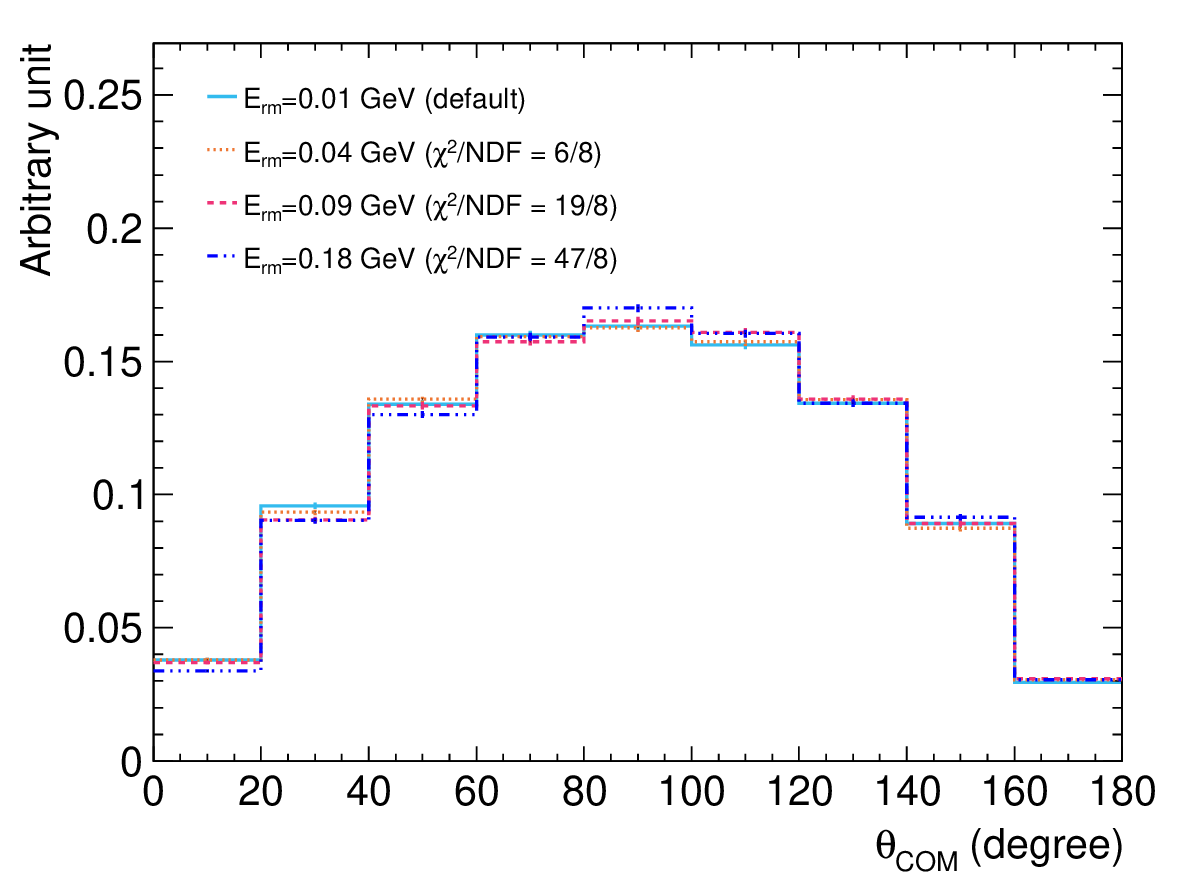}\\
        \caption{$\thetacom$}
        \label{subfig:ermc-comp-com}
    \end{subfigure}
    \begin{subfigure}[ht!]{\dbfigwid\textwidth}
        \centering
        \includegraphics[width=\textwidth]{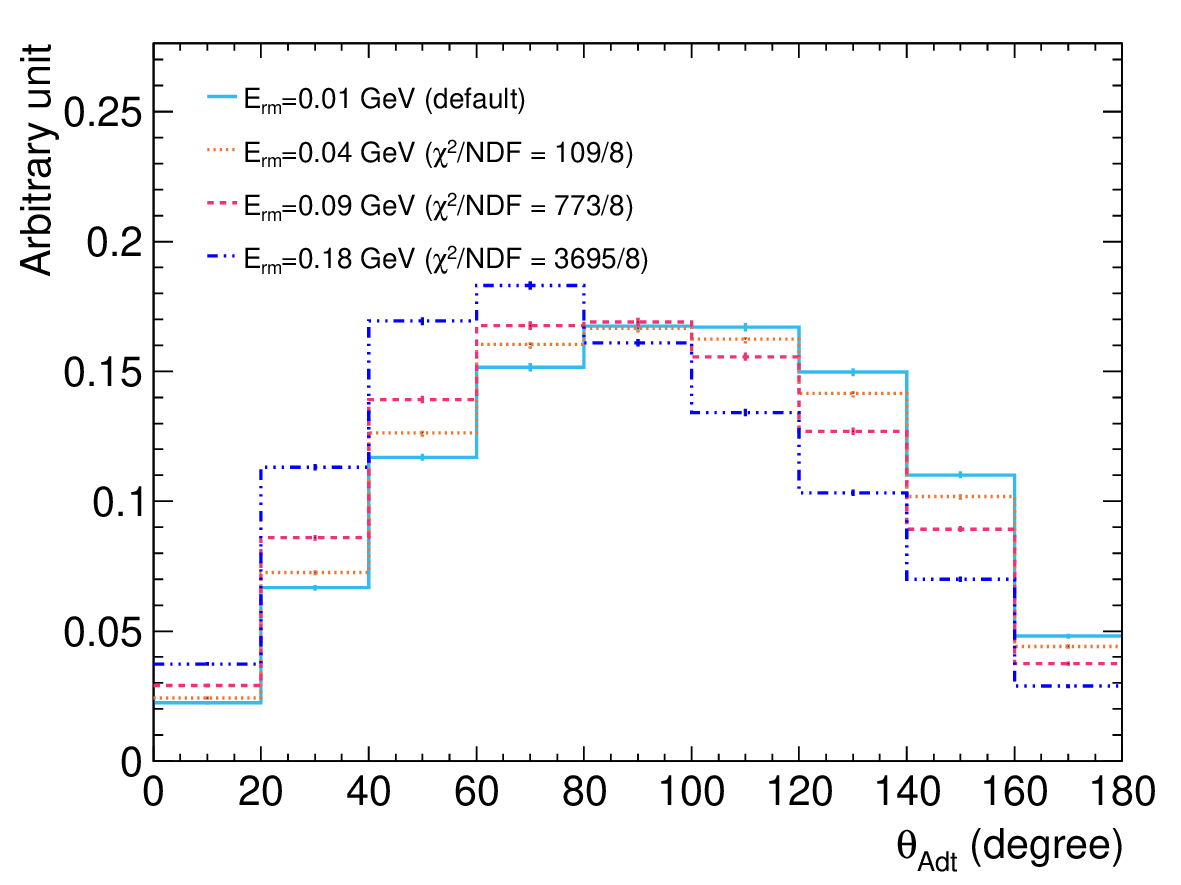}
        \caption{$\thetaadt$}
        \label{subfig:ermc-comp-adt}
    \end{subfigure}
    \caption{Area normalized comparisons for different removal energies, $E_{\textrm{rm}}$, with the T2K flux on carbon. The nominal tune is \gz.}
    \label{fig:ermc-comp}
\end{figure}

This observation could challenge the claimed IS insensitivity of $\thetacom$, as the agreement shown in Fig.~\ref{subfig:10alfg-comp-t2k} might be due to the relatively minor influence of IS.
A further stress test was conducted by varying the removal energy of carbon across a wide range—even to unphysical levels—to assess its effect on the shapes of $\thetacom$ and $\thetaadt$.
In \genie~, the removal energy is a parameter that describes the energy required to liberate a nucleon from the nucleus.
For a given incoming neutrino, a higher removal energy results in less of the energy transferred to the nuclear reaching the final state hadronic system.
The results of this comparison are presented in Fig.~\ref{fig:ermc-comp}.
As shown in Fig.~\ref{subfig:ermc-comp-com}, variations in the removal energy ($E_{\textrm{rm}}$) have very limited impact on $\thetacom$. 
The $\thetacom$ distribution remains statistically compatible with the default removal energy of $10~\mev$, even when $E_{\textrm{rm}}$ is increased to $90~\mev$.
When $E_{\textrm{rm}}$ is further increased to $180~\mev$, the $\thetacom$ distribution begins to deviate from the default case, although the $\chindf$ remains relatively small at $47/8$.
In contrast, the $\thetaadt$ peak exhibits a noticeable shift accompanied by a gradual change in its shape as $E_{\textrm{rm}}$ varies.
For $\thetaadt$, the $\chindf$ rises to $109/8$ when $E_{\textrm{rm}}$ is increased to $40~\mev$—a deviation much larger than that observed for $\thetacom$—and further increases to $3695/8$ when $E_{\textrm{rm}}$ is raised to $180~\mev$, which is unphysically high.
These results confirm that $\thetacom$ is robust against IS effects to a large extent, whereas $\thetaadt$ is more susceptible, as predicted.
With the strong robustness of $\thetacom$ against IS effects demonstrated relative to $\thetaadt$, the subsequent investigation will focus solely on $\thetacom$.

\begin{figure}
    \begin{subfigure}[b]{\dbfigwid\textwidth}
        \centering
        \includegraphics[width=\textwidth]{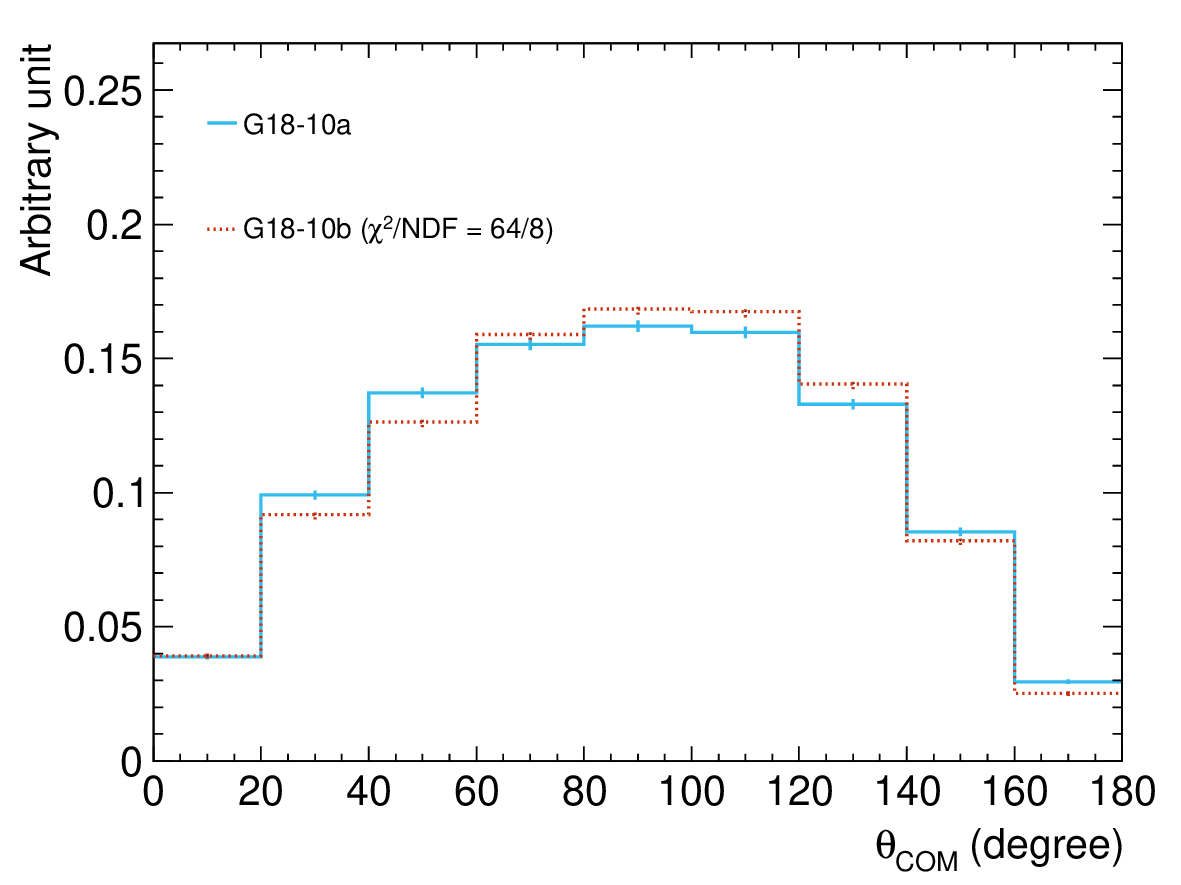}
        \caption{\geta~ vs \getb}
        \label{subfig:10a10b-comp-t2k}
    \end{subfigure}
    \begin{subfigure}[b]{\dbfigwid\textwidth}
        \centering
        \includegraphics[width=\textwidth]{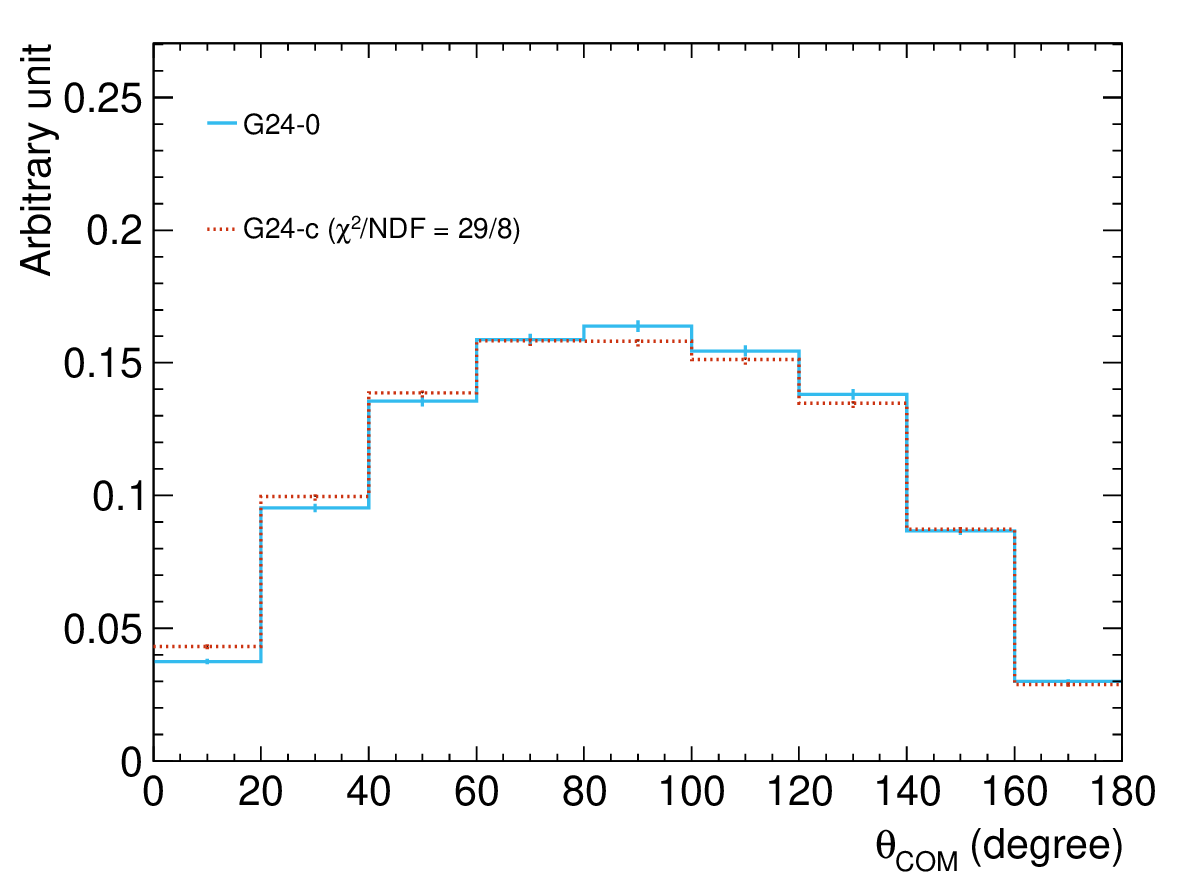}
        \caption{\gz~ vs \gc}
        \label{subfig:g240c-comp-t2k}
    \end{subfigure}
    \caption{Area normalized comparisons for different FSI models, \ref{subfig:10a10b-comp-t2k} for \geta~ (hA) and \getb~ (hN) and \ref{subfig:g240c-comp-t2k} for \gz~ (hA) and \gc~ (tuned hA) with the T2K flux on carbon.}
    \label{fig:fsi-comp}
\end{figure}

Another important property of $\thetacom$ to verify is its sensitivity to FSI effects.
Similar to the investigation of IS effects, the $\thetacom$ distributions for different FSI models are shown in Fig.~\ref{fig:fsi-comp}.
There are two commonly used FSI models in \genie, namely hA and hN.
Thus, two \genie~ tunes—\geta~ (hA) and \getb~ (hN), which differ only in FSI modeling—are compared in Fig.~\ref{subfig:10a10b-comp-t2k}.
The $\thetacom$ distributions for these two tunes are statistically incompatible, with a large $\chindf$ value of $64/8$, a difference that is significantly more pronounced than the fluctuation induced by changing the IS model from LFG to CFG, as shown in Fig.~\ref{subfig:10alfg-comp-t2k}.
To further verify the sensitivity of $\thetacom$ to FSI effects, a comparison between \gz~ (hA) and \gc~ (tuned hA) is presented in Fig.~\ref{subfig:g240c-comp-t2k}.
In \gc~\cite{GENIE:2024ufm}, the hA model has been tuned to maintain good agreement with the T2K TKI $\numuccopiop$ data~\cite{T2K:2021naz}, so the considerable difference—$\chindf=29/8$—between the two distributions in Fig.~\ref{subfig:g240c-comp-t2k} further demonstrates the sensitivity of $\thetacom$ to FSI effects.

\begin{figure}
    \centering
    \begin{subfigure}[b]{\dbfigwid\textwidth}
        \centering
        \includegraphics[width=\textwidth]{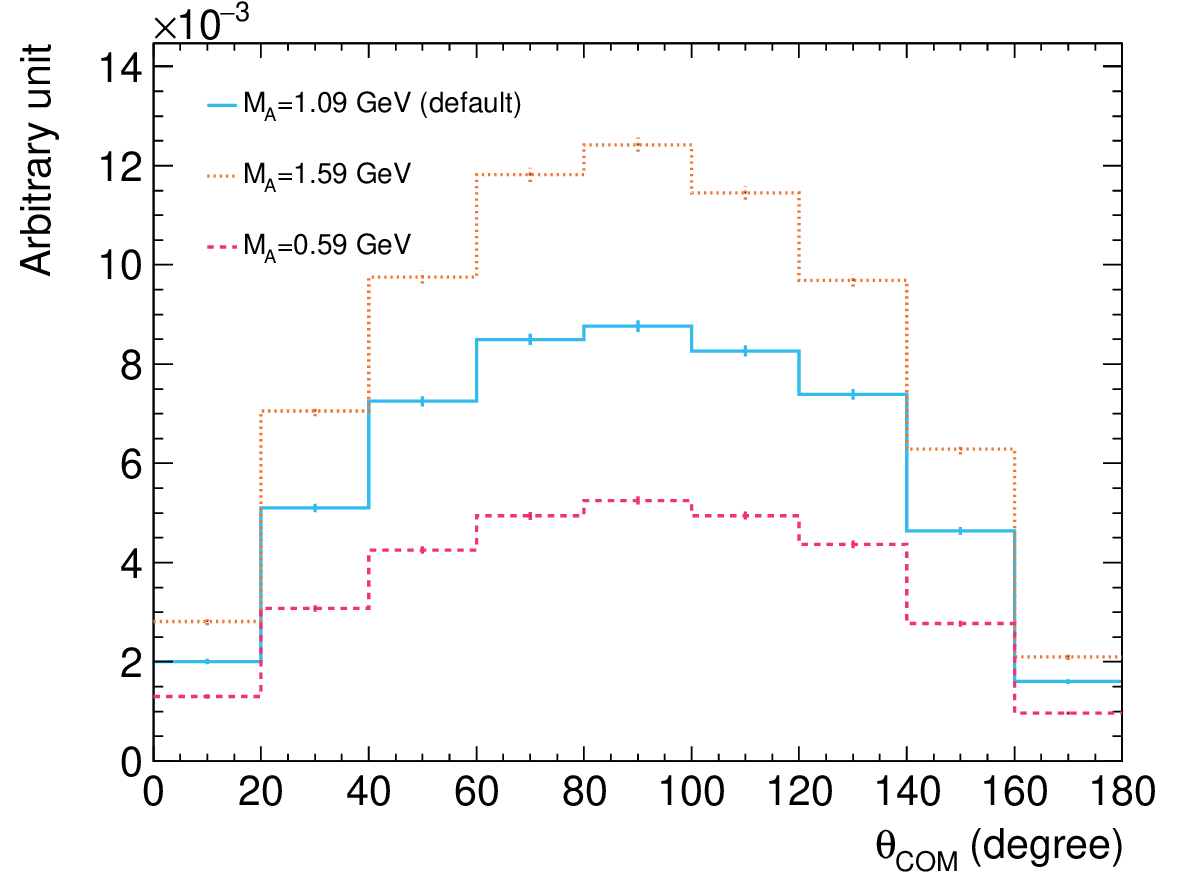}
        \caption{Cross-section (per nucleus) normalized}
        \label{subfig:ma-comp-xsec}
    \end{subfigure}
    \begin{subfigure}[b]{\dbfigwid\textwidth}
        \centering
        \includegraphics[width=\textwidth]{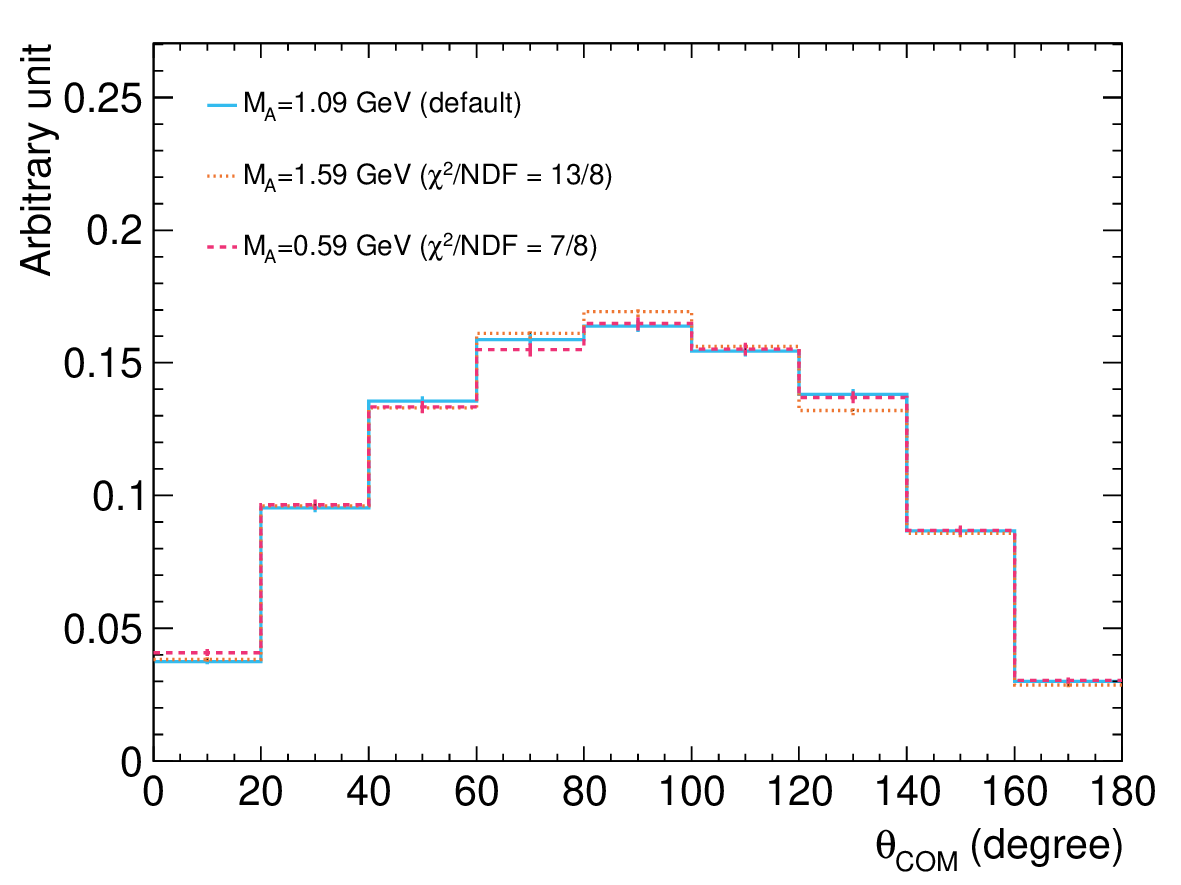}
        \caption{Area normalized}
        \label{subfig:ma-comp-area}
    \end{subfigure}
    \caption{Comparisons for different $\MA$ values with the T2K flux on carbon. The nominal tune is \gz.}
    \label{fig:ma-comp}
\end{figure}

The two major advantages of $\thetacom$ have been demonstrated, namely its robustness against IS effects and its sensitivity to FSI effects.
The next step is to investigate the impact of other factors in neutrino–nucleus interactions on $\thetacom$.
Since $\thetacom$ is calculated from the decay products of resonances, it can be affected by changes in RES modeling, which dictate the production and decay of these resonances.
As a proxy for investigating the effect of a change in the RES model on $\thetacom$, the axial mass parameter, $\MA$, in the \gz~ tune was varied to an exaggerated—and most likely unphysical—extent, as shown in Fig.~\ref{fig:ma-comp}.
As anticipated, Fig.~\ref{subfig:ma-comp-xsec} demonstrates that varying $\MA$ alters the cross-section, while the shape of $\thetacom$ remains almost unchanged—as confirmed by the small $\chindf$ values in Fig.~\ref{subfig:ma-comp-area}.
This is likely because $\deltapp$ is the dominant resonance for the T2K flux, and changes in $\MA$ do not appreciably alter its decay properties—thereby leaving the $\thetacom$ distribution largely unaffected.

\begin{figure}
    \centering
    \begin{subfigure}[b]{\qufigwid\textwidth}
        \centering
        \includegraphics[width=\textwidth]{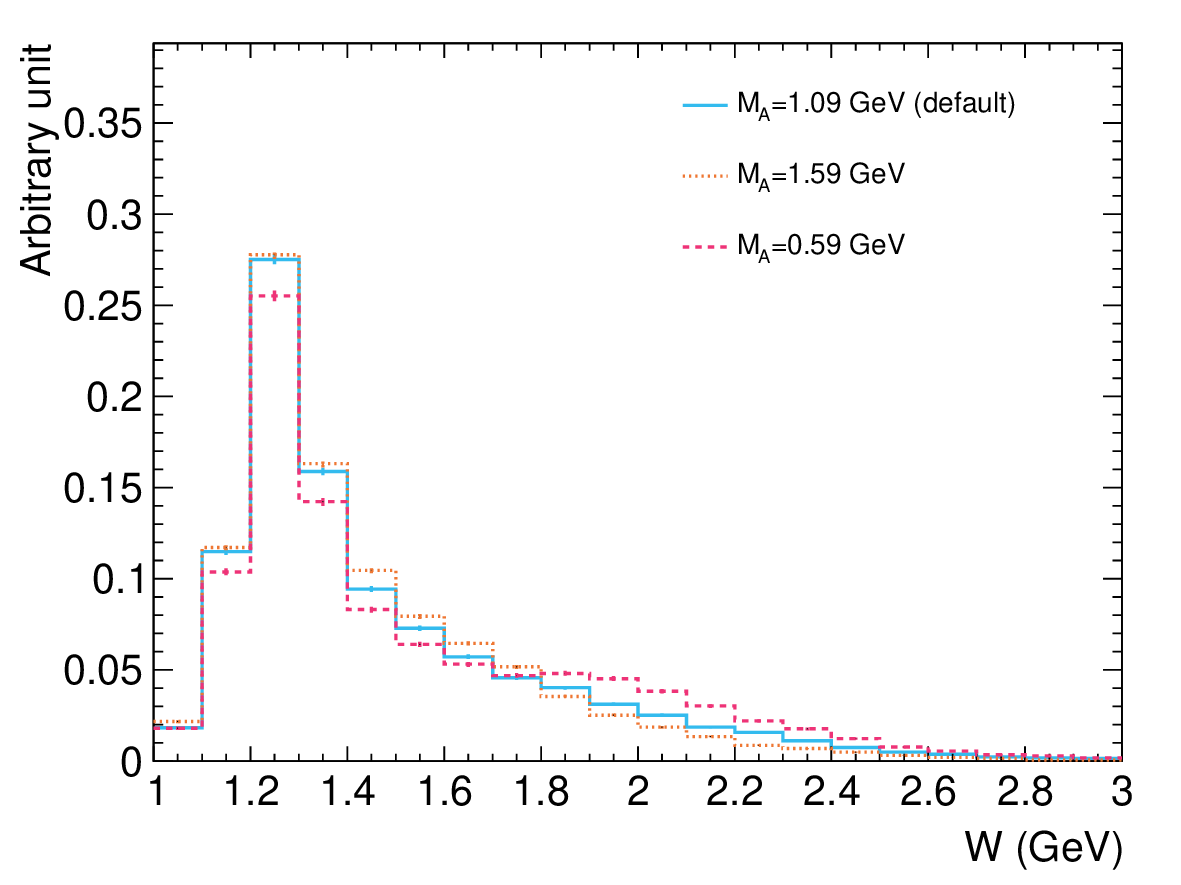}
        \caption{$\numuccopiop$}
        \label{subfig:ma-comp-minerva-w}
    \end{subfigure}
    \begin{subfigure}[b]{\qufigwid\textwidth}
        \centering
        \includegraphics[width=\textwidth]{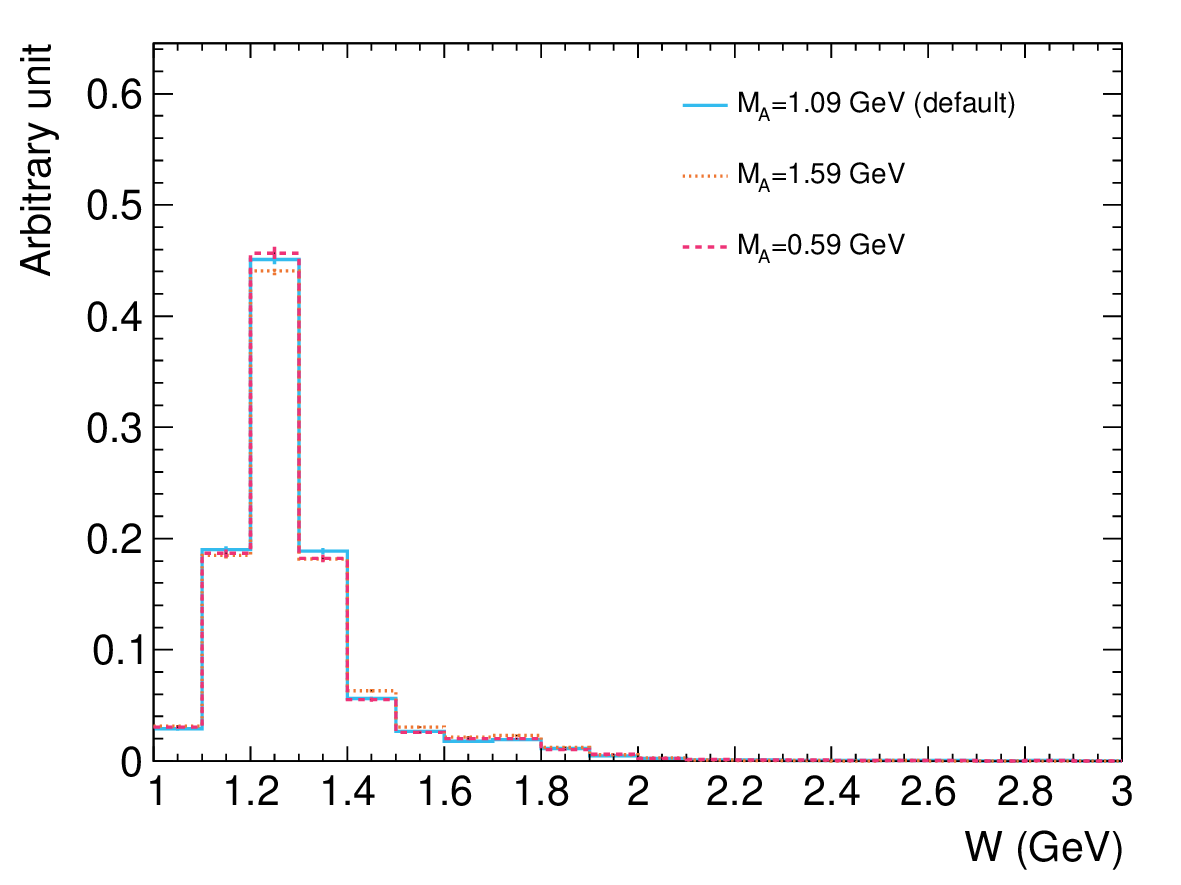}
        \caption{$\numuccopiop$ with $\ecom<1330~\mev$}
        \label{subfig:ma-comp-minerva-wcut-w}
    \end{subfigure}
    \\
    \begin{subfigure}[b]{\qufigwid\textwidth}
        \centering
        \includegraphics[width=\textwidth]{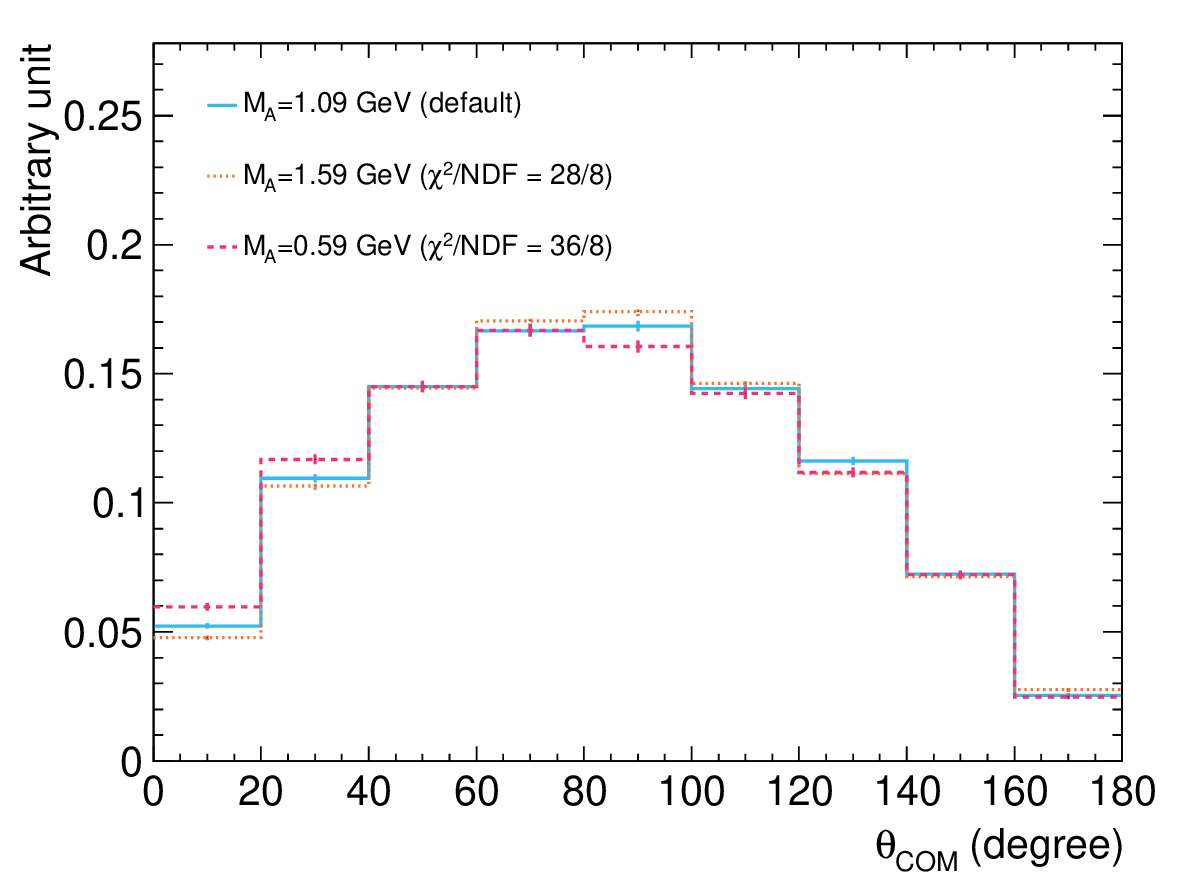}
        \caption{$\numuccopiop$}
        \label{subfig:ma-comp-minerva}
    \end{subfigure}
    \begin{subfigure}[b]{\qufigwid\textwidth}
        \centering
        \includegraphics[width=\textwidth]{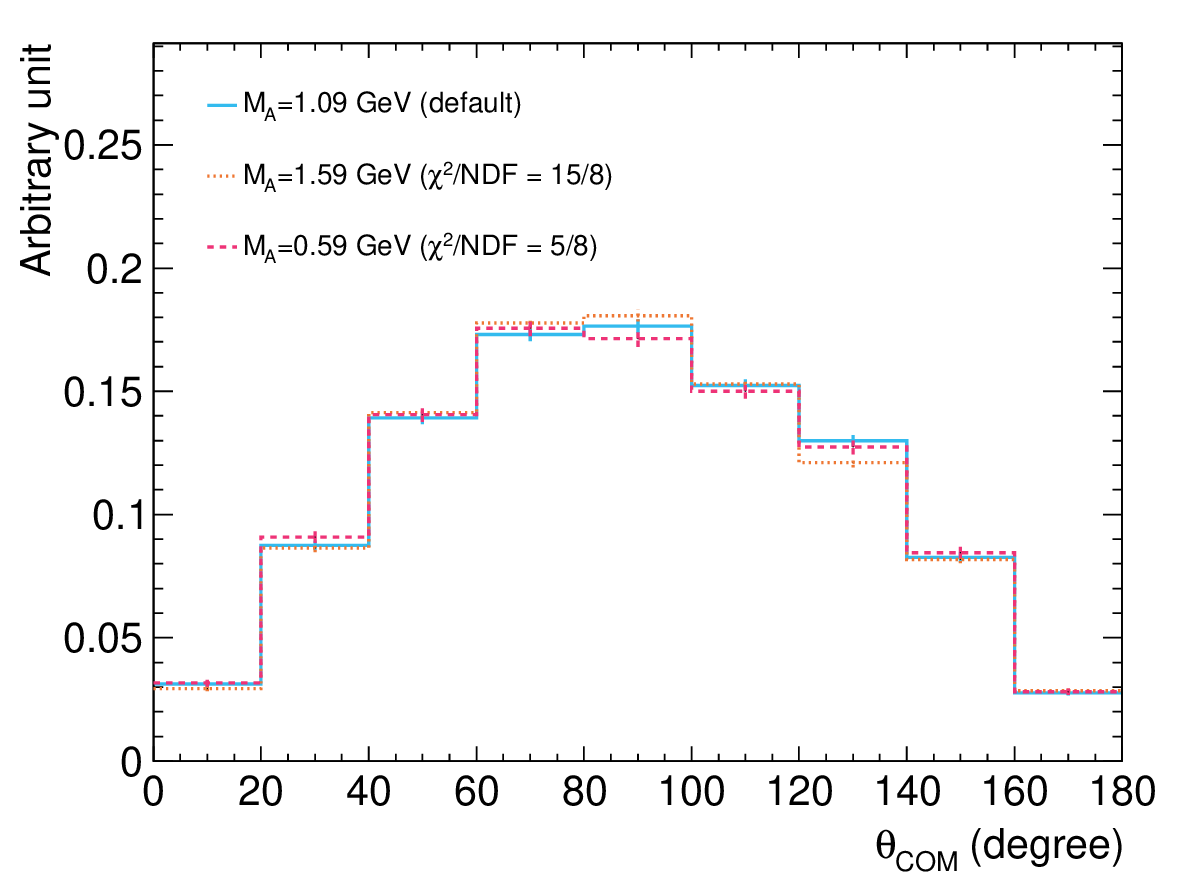}
        \caption{$\numuccopiop$ with $\ecom<1330~\mev$}
        \label{subfig:ma-comp-minerva-wcut}
    \end{subfigure}
    \caption{Area normalized comparisons of $W$ and $\thetacom$ for different $\MA$ values with the MINERvA flux on the MINERvA target. The nominal tune is \gz.}
    \label{fig:ma-comp-minerva}
\end{figure}

To complement this investigation, a similar comparison conducted using the MINERvA flux on the MINERvA target is shown in Fig.~\ref{fig:ma-comp-minerva}. 
Since the MINERvA flux is more energetic, higher resonances also make a significant contribution.
Changes in $\MA$ can affect the resonances differently, as demonstrated by the $W$ distribution, which represents the invariant mass of the pion-nucleon system estimated from leptonic kinematics. 
The derivation of $W$ is readily available in the literature, such as in Ref.~\cite{Paschos:2003qr}.
As evident in Fig.~\ref{subfig:ma-comp-minerva-w}, when $\MA$ decreases to $0.59~\gev$, the relative contribution of resonances above $1.8~\gev$ increases, while the contribution of those below decreases appreciably.
In contrast, when $\MA$ increases to $1.59~\gev$, the relative contribution of resonances above $1.8~\gev$ decreases, while the contribution of those below either increases or remains relatively unchanged.
Consequently, these changes in $\MA$ lead to considerable differences in the $\thetacom$ distributions, with $\chindf=28/8$ for $\MA=1.59~\gev$ and $\chindf=44/8$ for $\MA=0.59~\gev$, as shown in Fig.~\ref{subfig:ma-comp-minerva}.
To minimize the influence of higher resonances, a practical cut, $\ecom<1330~\mev$, aimed at selecting $\deltapp$ events, is imposed on the selection in Fig.~\ref{subfig:ma-comp-minerva-wcut}.
As anticipated, the higher resonances in the $W$ distribution have largely disappeared in Fig.~\ref{subfig:ma-comp-minerva-wcut-w}, and the $\thetacom$ distributions for different $\MA$ values in Fig.~\ref{subfig:ma-comp-minerva} become statistically compatible.
This suggests that for an energetic flux such as the MINERvA flux—which produces multiple resonances—the $\thetacom$ distribution can still be used to study FSI with minimal impact from RES modeling by imposing an $\ecom$ cut to select only $\deltapp$ events.

\begin{figure}
    \centering
    \begin{subfigure}[b]{\dbfigwid\textwidth}
        \centering
        \includegraphics[width=\textwidth]{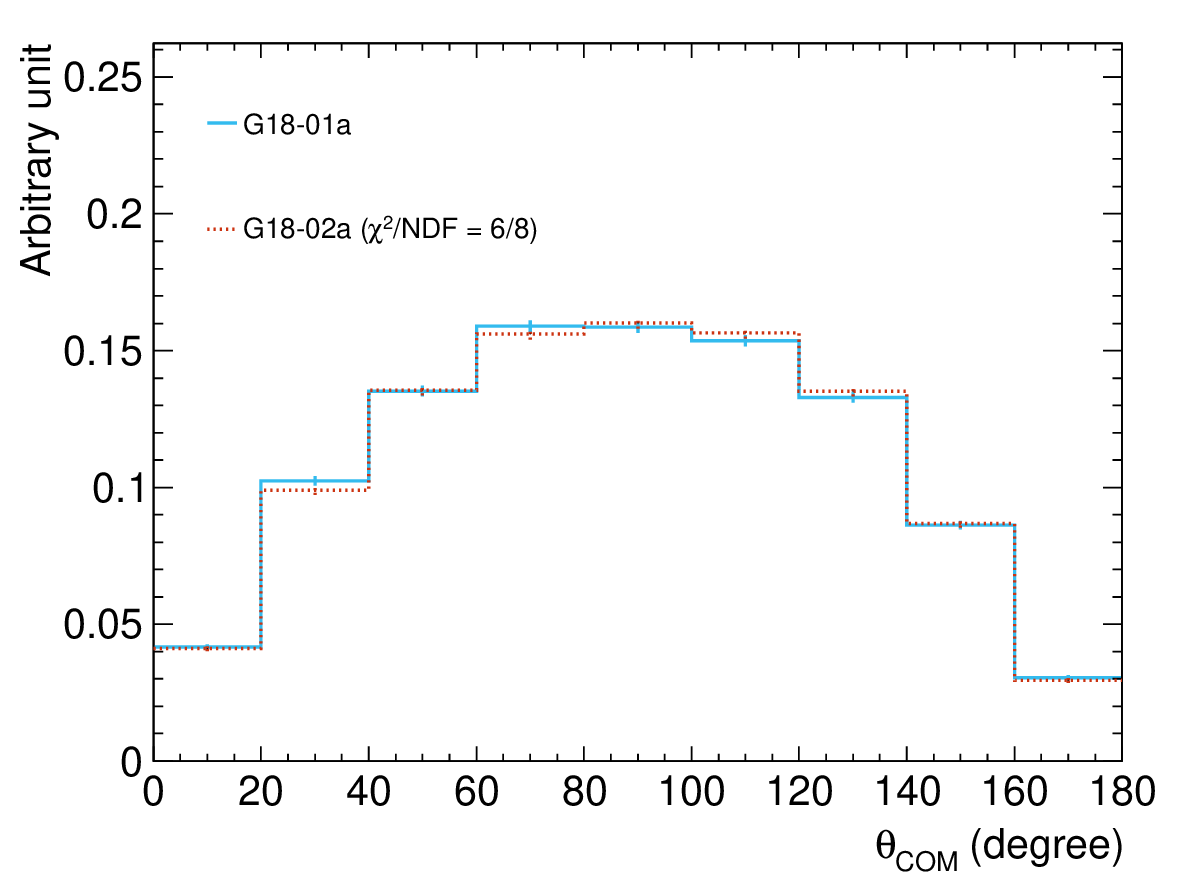}
        \caption{T2K}
        \label{subfig:0102a-comp-t2k}
    \end{subfigure}
    \begin{subfigure}[b]{\dbfigwid\textwidth}
        \centering
        \includegraphics[width=\textwidth]{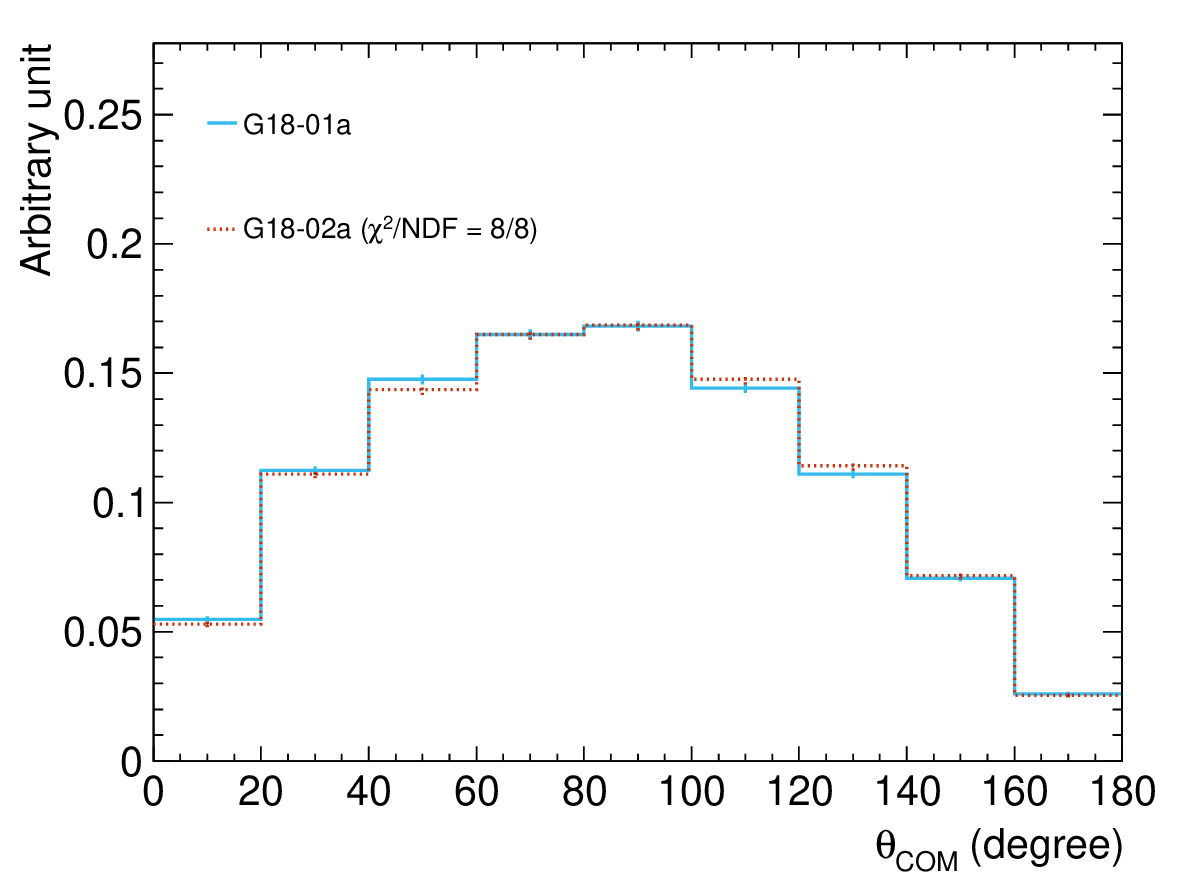}
        \caption{MINERvA}
        \label{subfig:0102a-comp-minerva}
    \end{subfigure}
    \caption{Area normalized comparisons for different RES models with the T2K (\ref{subfig:0102a-comp-t2k}) and MINERvA (\ref{subfig:0102a-comp-minerva}) fluxes using \geoa~ and \getwoa.
    Both select $\numuccopiop$ events, but the MINERvA sample has a additional cut of $\ecom<1330~\mev$.}
    \label{fig:0102a-comp}
\end{figure}

Besides varying a parameter in the RES model, a more aggressive test is to compare two different RES models directly.
This comparison is achieved by contrasting two \genie~ tunes—\geoa~ (employing the RS model) and \getwoa~ (employing the BS model)—as shown in Fig.~\ref{fig:0102a-comp}.
The results using the T2K flux and the MINERvA flux are presented in Fig.~\ref{subfig:0102a-comp-t2k} and Fig.~\ref{subfig:0102a-comp-minerva}, respectively.
Both $\chindf$ values are low—specifically, $6/8$ and $8/8$—suggesting that this change in the RES model is far less extreme than the unphysically large modification of $\MA$ shown in Fig.~\ref{fig:ma-comp} and that such a reasonable change in RES modeling does not lead to a noticeable alteration of the $\thetacom$ distribution.

On one hand, restricting the analysis to $\deltapp$ events—by imposing an $\ecom$ cut—can minimize the impact of RES modeling on $\thetacom$ even for an energetic beam like the MINERvA flux.
On the other hand, this behavior implies that $\thetacom$ is sensitive to the onset of higher resonances.
Given that the robustness of $\thetacom$ against both IS effects and considerable changes of RES modeling have been demonstrated, the difference between the $\thetacom$ distributions with and without the $\ecom$ cut (as shown in Fig.~\ref{fig:ma-comp-minerva}) is most likely attributable to the onset of higher resonances.
This advocates employing $\thetacom$ to study both the production of higher resonances and the correlations among their production.
Since the BS model in \genie~ does not implement resonance correlations, this potential application of $\thetacom$ is reserved for future studies—when more advanced models, such as the MK model~\cite{Kabirnezhad:2017jmf,Kabirnezhad:2020wtp,Kabirnezhad:2022znc}, are reliably implemented and validated.

Besides its appealing sensitivity and robustness, $\thetacom$ has the practical advantage that its reconstruction does not require determining the kinematics of the incoming neutrino.
This decouples the reconstruction of $\thetacom$ from the systematic uncertainties associated with neutrino energy reconstruction—which are among the largest in neutrino measurements~\cite{T2K:2019yqu,T2K:2021naz,MicroBooNECollaboration:2024gvg,NOvA:2023uxq,MINERvA:2022djk}.
However, the reconstruction of $\thetacom$ still depends on the accurate reconstruction of proton kinematics, which is itself a significant source of systematic uncertainty.
The overall impact on precision for $\thetacom$ measurements will be investigated in future studies for specific experiments.

\begin{figure}[ht!]
    \centering
    \begin{subfigure}[ht!]{\dbfigwid\textwidth}
        \centering
        \includegraphics[width=\textwidth]{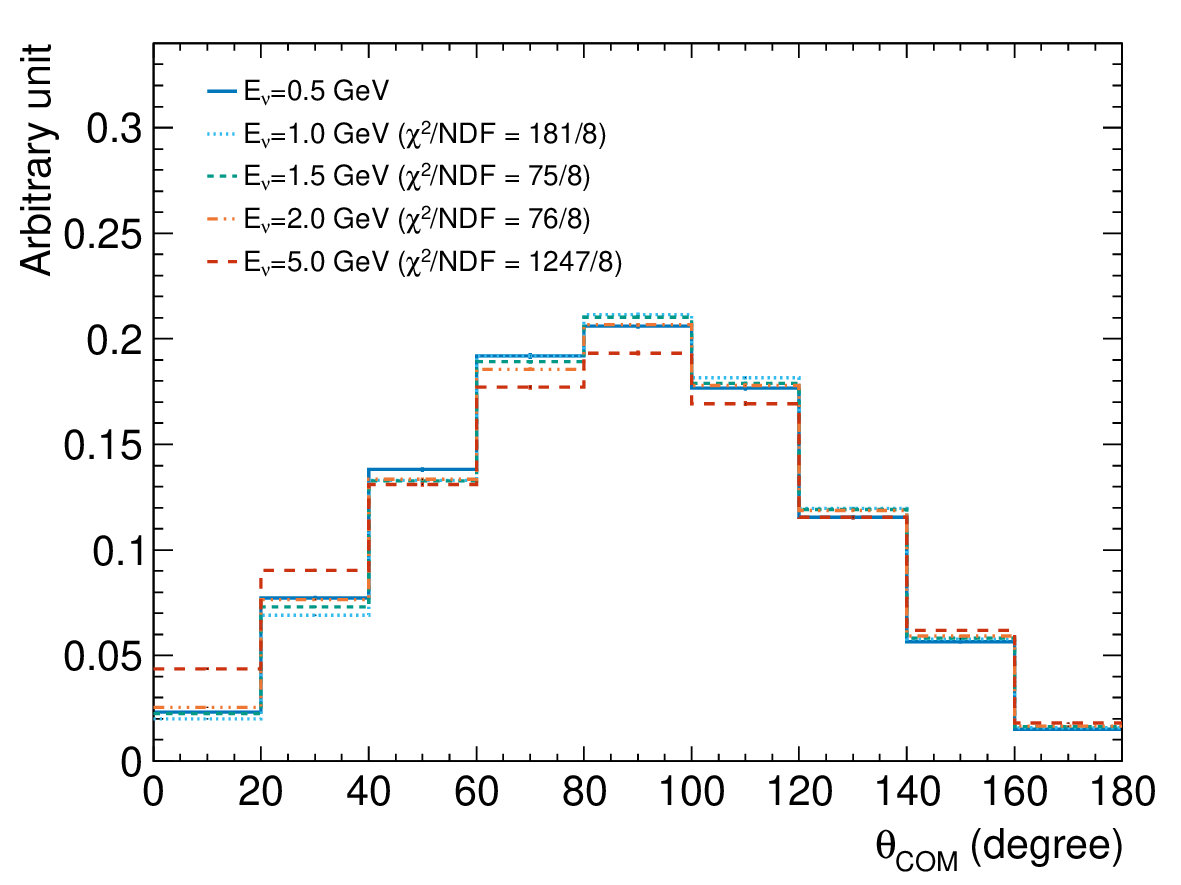}
        \caption{$\numuccopiop$}
        \label{subfig:enu-comp-cc1pi1p}
    \end{subfigure}
    \begin{subfigure}[ht!]{\dbfigwid\textwidth}
        \centering
        \includegraphics[width=\textwidth]{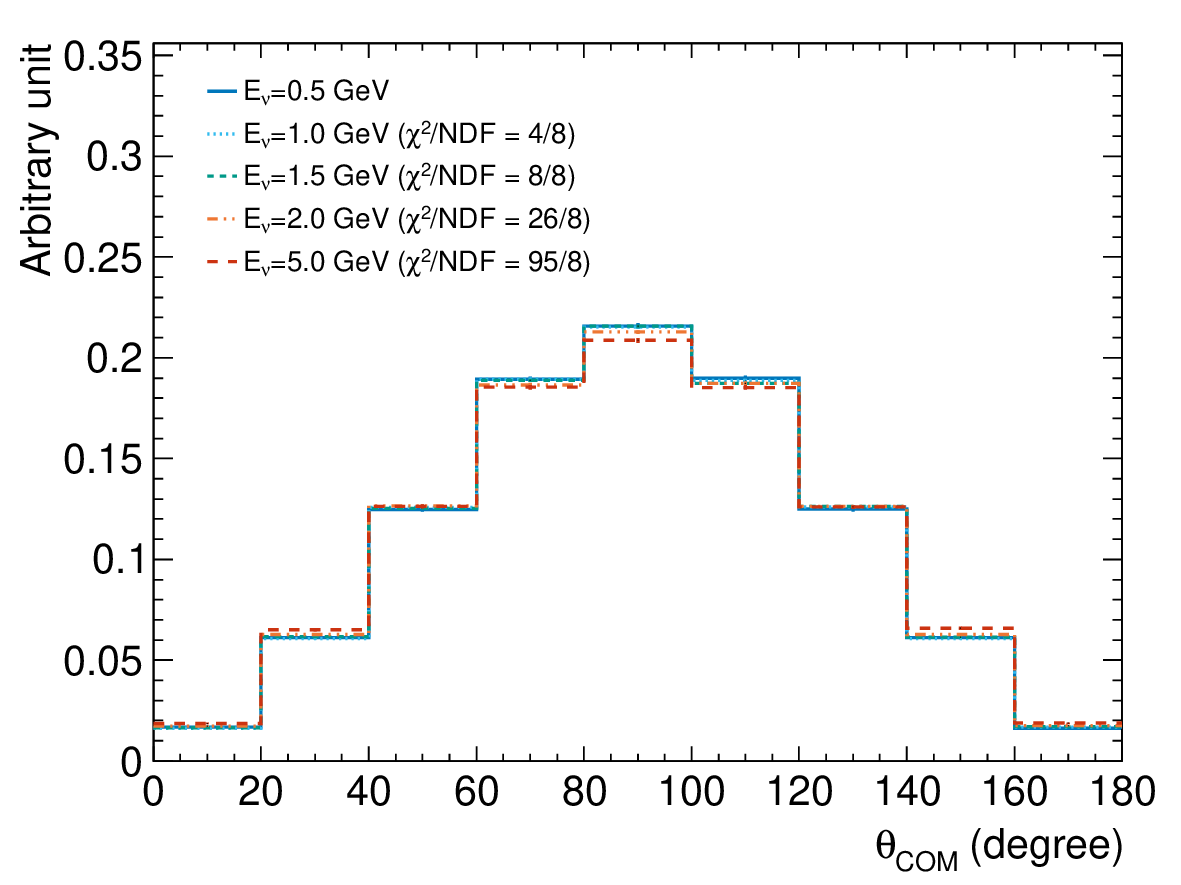}
        \caption{$\deltapp$ only}
        \label{subfig:enu-comp-dpp}
    \end{subfigure}
    \caption{Area normalized comparisons for different $\enu$ fluxes on hydrogen for $\thetacom$. The nominal tune is \gz.}
    \label{fig:enu-comp-h}
\end{figure}

To assess the impact of $\enu$ on $\thetacom$, mono-energy neutrino beams are used to simulate $\nu$-H events using tune \gz, as shown in Fig.~\ref{fig:enu-comp-h}.
Hydrogen is chosen as the target to isolate the impact solely due to $\enu$, while complications arising from FSI will be discussed later.

In Fig.~\ref{subfig:enu-comp-cc1pi1p}, although the shapes of the $\thetacom$ distributions appear largely consistent for $\enu$ values between $0.5~\gev$ and $2.0~\gev$, they are statistically incompatible—with $\chindf$ values on the order of $100/8$. 
This inconsistency is likely due to the onset of higher doubly positive resonances. 
While $\thetacom$ for a single resonance is largely independent of the production mechanism—unless a strong correlation exists between production and decay—it will be affected when the relative contributions of different resonances vary due to their distinct decay properties. 
If the analysis is restricted to $\deltapp$ events (based on true information), as in Fig.~\ref{subfig:enu-comp-dpp}, it is encouraging to observe that all $\enu$ distributions—except for $\enu=5.0~\gev$—become compatible with much smaller $\chindf$ values, thereby confirming the independence of $\thetacom$ from $\enu$ for a single resonance.

In the case of $\enu=5.0~\gev$, much more energetic $\deltapp$ resonances are produced, which enhance the influence of $\deltapp$ kinematics on their decay and tend to favor either very small or very large pion angles.
Since such highly energetic neutrinos contribute only marginally to actual cross‐section measurements, this effect is not expected to pose a significant challenge for the practical application of $\thetacom$ in neutrino analyses.
A comprehensive explanation of this observation is beyond the scope of this work; instead, a more detailed investigation of the correlation between resonance production and decay is warranted in future studies.

\begin{figure}[ht!]
    \centering
    \begin{subfigure}[ht!]{\dbfigwid\textwidth}
        \centering
        \includegraphics[width=\textwidth]{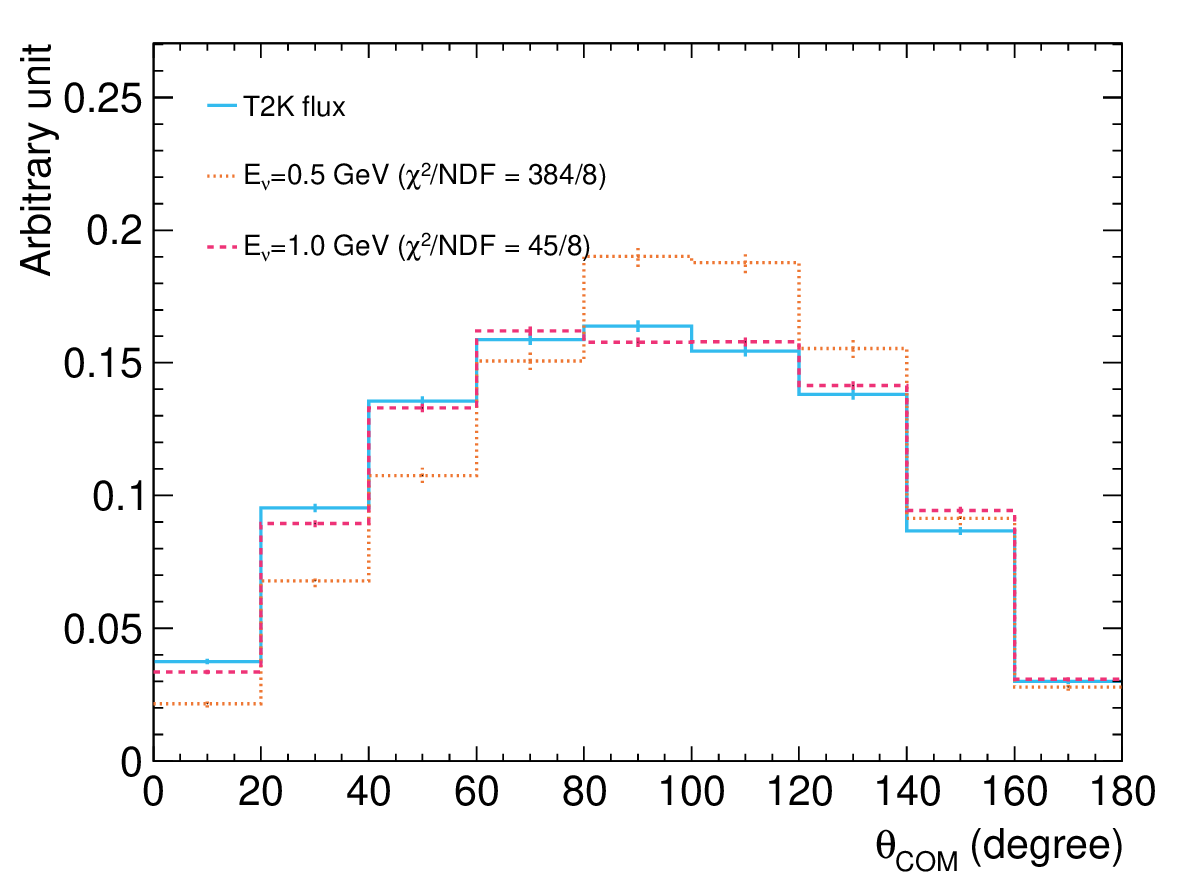}
        \caption{$\numuccopiop$}
        \label{subfig:enu-comp-cc1pi1p-c}
    \end{subfigure}
    \begin{subfigure}[ht!]{\dbfigwid\textwidth}
        \centering
        \includegraphics[width=\textwidth]{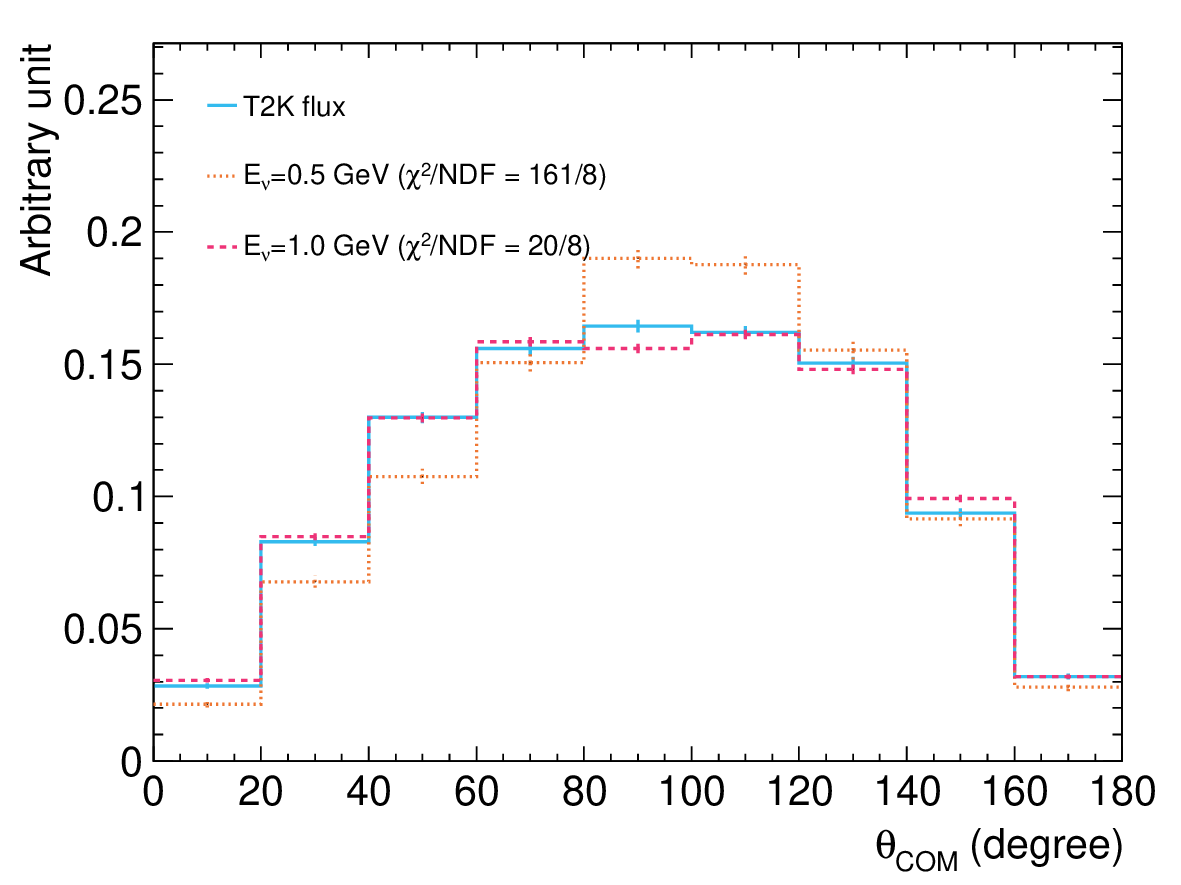}
        \caption{$\numuccopiop$ with $\ecom<1330~\mev$}
        \label{subfig:enu-comp-dpp-c}
    \end{subfigure}
    \caption{Area normalized comparisons for different $\enu$ fluxes and the T2K flux on carbon for $\thetacom$. The nominal tune is \gz.}
    \label{fig:enu-comp-c}
\end{figure}

In the resonance rest frame, the pion decay angle is an intrinsic property of the resonance and should be largely independent of its kinematics—and consequently, of $\enu$.
However, in the lab frame the boost experienced by the decay products results in different hadronic kinematic distributions.
Since FSI depends on the hadronic kinematics, the measured kinematics—and hence $\thetacom$—are indirectly affected by $\enu$.
Therefore, although the reconstruction of $\thetacom$ for an individual event is independent of $\enu$, the overall $\thetacom$ distribution may vary due to changes in the neutrino energy spectrum when FSI is present.
To assess this impact in practice, mono-energetic neutrino beams are used to simulate $\nu$-C events using tune \gz, and the results are compared to those obtained with the T2K flux, as shown in Fig.~\ref{fig:enu-comp-c}.

As expected, variations in hadronic kinematics caused by different $\enu$ values lead to noticeable changes in the $\thetacom$ distribution, as illustrated in Fig.~\ref{subfig:enu-comp-cc1pi1p-c}.
This effect is more pronounced for $\enu=0.5~\gev$, where low-momentum pions—the decay products of low-momentum $\deltapp$ resonances resulting from low-energy neutrinos—are more likely to be absorbed through FSI.
Thus, even with an $\ecom < 1330~\mev$ cut imposed, the $\thetacom$ distribution for $\enu=0.5~\gev$ remains incompatible with that for the T2K flux, whereas the $\enu=1.0~\gev$ distribution becomes relatively compatible with the T2K flux, exhibiting a $\chindf$ value of $20/8$.
This suggests that while $\thetacom$ can exhibit robustness against $\enu$ variations when $\ecom$ cuts are applied to isolate a specific resonance, caution must be exercised regarding the influence of $\enu$ on the kinematics of the resonance. 
This effect becomes particularly pronounced at low energies, where the decay products are more susceptible to FSI. 
In this study, the $\enu$ robustness has been evaluated for the T2K flux. 
However, it is anticipated that a comparable degree of robustness could be achieved for other fluxes by employing flux-specific cuts—a possibility that warrants further investigation.

\begin{figure}
    \includegraphics[width=\scfigwid\textwidth]{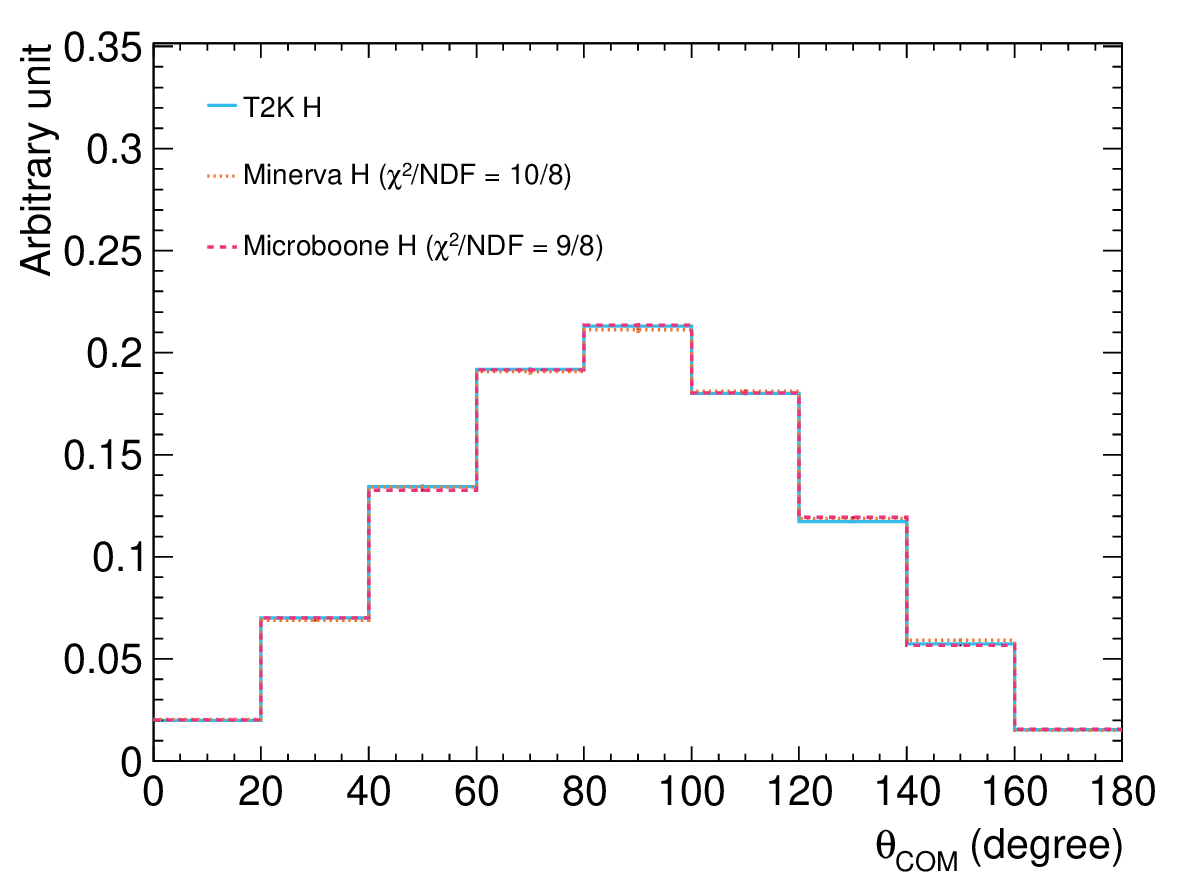}
    \caption{Area normalized comparisons for different fluxes on hydrogen with the $\ecom<1330~\mev$ cut imposed for MINERvA. The nominal tune is \gz.}
    \label{fig:flux-comp}
\end{figure}

The observed robustness of $\thetacom$ against $\enu$ variations in $\nu$-H events—across a wide range of $\enu$ with an $\ecom < 1330~\mev$ cut, as seen in Fig.~\ref{fig:enu-comp-h}—opens up a new avenue for cross-experiment comparisons.
As depicted in Fig.~\ref{fig:flux-comp}, the $\thetacom$ distributions for the T2K, MINERvA, and MicroBooNE~\footnote{There is no hydrogen in the target for MicroBooNE, but such a simulation is possible and plotted for reference only.} fluxes on a hydrogen target using tune \gz are highly statistically compatible, with $\chindf$ values of $10/8$ and $9/8$, respectively.
The advantage of a $\nu$-H selection is that it avoids the complications introduced by FSI affecting the hadronic kinematics in the lab frame, as illustrated in Fig.~\ref{fig:enu-comp-c}.
Consequently, despite differing energy profiles, the neutrino fluxes from these experiments yield nearly identical $\thetacom$ distribution shapes.
In practice, MINERvA~\cite{MINERvA:2023avz} has already conducted a $\nu$-H selection, and the T2K collaboration is planning similar analyses~\cite{Lu:2015hea}, thereby demonstrating the feasibility of such cross-experiment comparisons in the future.
Given a high‐purity hydrogen sample, the $\thetacom$ cross section can be directly compared across different experiments.

\begin{figure}[ht!]
    % \begin{subfigure}[ht!]{\scfigwid\textwidth}
        \centering
        \includegraphics[width=\scfigwid\textwidth]{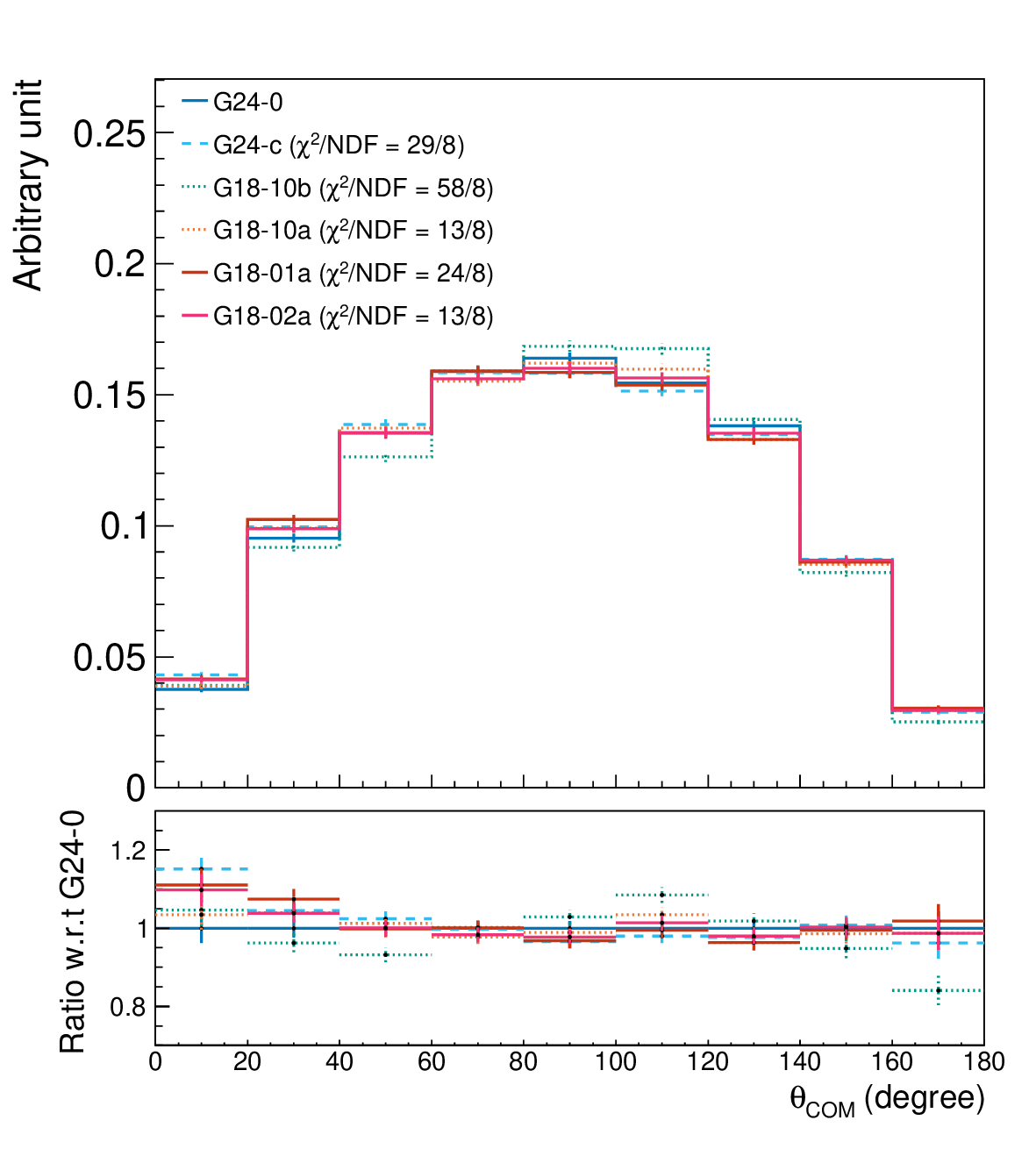}
        % \caption{}
        % \label{subfig:mod-comp-ratio}
    % \end{subfigure}
    \caption{$\thetacom$ distribution comparisons using the T2K flux on carbon with multiple \genie~ configuations. }
    \label{fig:mod-comp}
\end{figure}

Thus far, the discussion has primarily focused on the insensitivity of $\thetacom$ to variations in one aspect (excluding FSI) of the complex neutrino–nucleus interactions.
To further demonstrate the robustness of $\thetacom$, a comparison among different \genie~ tunes is presented in Fig.~\ref{fig:mod-comp}, in which multiple components of the neutrino–nucleus interactions vary concurrently.

Notably, the configurations differing in their FSI model—\gc~ (tuned hA) and \getb~ (hN)—exhibit the largest $\chindf$ values, $29/8$ and $58/8$, respectively.
\gc differs from \gz only in its FSI model, as discussed in Fig.~\ref{subfig:g240c-comp-t2k}.
In addition to differing in the FSI model, \getb~ also differs from \gz~ in its IS and 2p2h models.
In contrast, \geta—which shares the same FSI model as \gz~ but differs in IS and 2p2h models in a manner similar to \getb—exhibits a much smaller $\chindf$ value of $13/8$, providing strong evidence that FSI is the primary reason for the large deviation observed in \getb~ relative to \gz.

Furthermore, \geoa~ and \getwoa~ differ from \gz~ in additional model components, yet both exhibit relatively small $\chindf$ values.
In particular, \getwoa~ is statistically compatible with \gz—with a $\chi^2/\textrm{NDF}$ of $13/8$—despite differences in its IS, QE, and 2p2h models.
While a change in the RES model alone does not impact $\thetacom$, as illustrated by the comparison between \geoa~ and \getwoa~ in Fig.~\ref{fig:0102a-comp}, the combined modifications in QE, IS, and 2p2h models cause \geoa~ to deviate further from \gz, yielding a $\chindf$ of $24/8$.

This comparison between different \genie~ tunes demonstrates that $\thetacom$ is most sensitive to changes in FSI while remaining robust against variations in other model components.
A change in the RES model combined with modifications in other components can result in a noticeable alteration of the $\thetacom$ distribution, whereas a change in the RES model alone does not appear to have a significant impact.

\section{Discussion}
\label{sec:dis}
The simulation studies in this work demonstrate that the COM angle is a novel variable with multiple advantages, and its measurement will be valuable for a focused study on FSI with minimal influence from other aspects of neutrino–nucleus interactions.
Models constrained by $\thetacom$ measurements—e.g., following the methodology in Ref.~\cite{GENIE:2021zuu}—will remain relevant even if other aspects of neutrino–nucleus interactions, in particular the IS model, are updated in the future.
This is an advantage that variables—such as $\thetaadt$, which depend on multiple processes of neutrino–nucleus interactions—do not possess.
Furthermore, the analyses presented are based on true information without realistic fluctuations, meaning that challenges related to reconstruction—particularly those involving neutrino energy for the Adler angle—are not accounted for, whereas $\thetacom$ exhibits minimal dependence on neutrino energy reconstruction for T2K, as shown in Fig.~\ref{subfig:enu-comp-dpp-c}.
Therefore, the full potential of the COM angle will be better realized in a cross-section measurement that accounts for these complexities.

Although the focus of the $\thetacom$ analysis is on nuclear effects, it is inherently affected by resonance interaction modeling.
Fortunately for T2K, resonance production is dominated by the $\deltapp$, thereby rendering the $\deltapp$ measurement at T2K resistant to changes in RES modelling, as shown in Fig.~\ref{fig:ma-comp} and Fig.~\ref{subfig:0102a-comp-t2k}.
In contrast, the more energetic MINERvA flux produces multiple resonances, and the $\thetacom$ measurement is more sensitive to changes in RES modelling, as shown in Fig.~\ref{fig:ma-comp-minerva}.
This sensitivity to RES modeling can be managed or even exploited in different ways.

Firstly, imposing a cut on $\ecom$ to select $\deltapp$ events can reduce this sensitivity, as shown in Fig.~\ref{subfig:ma-comp-minerva-wcut} and Fig.~\ref{subfig:0102a-comp-minerva}, and under this condition, the $\thetacom$ measurement can be used for focused studies of FSI, similar to the T2K case.
Secondly, comparing the $\thetacom$ distributions with and without the $\ecom$ cut can be used to study the onset of higher resonances and the correlations among different resonances.

As noted in Sec.~\ref{sec:com}, the $\thetacom$ distribution is influenced by the production of higher resonances that decay via the same channels as $\deltapp$.
Due to the general difference between $\thetaR$ and $\thetapidel$, $\thetacom$ will start to deviate from $\thetapidel$ (smeared by FSI), as shown in Fig.~\ref{subfig:enu-comp-cc1pi1p}.
Although $\thetacom$ is inspired by $\thetapidel$, it is not restricted to the reconstruction or estimation of $\thetapidel$. 
Rather, by definition, $\thetacom$ is a superposition of all energetically accessible $\thetaR$, with $\thetapidel$ as the dominant component at low energy.
If all $\thetaR$ values are relatively well understood, a simple, theoretically motivated expectation of $\thetacom$ can be obtained by summing the contributions from all possible resonances according to their relative production ratios.

This approach is complicated further by the fact that even with a complete understanding of the FSI model, the predicted $\thetacom$ distribution may deviate from measurements due to correlations among different resonance production modes; that is, the overall $\thetacom$ is not simply a linear combination of the individual $\thetaR$ contributions.
Modeling of such correlations is currently absent in the BS model but is accounted for in the MK model~\cite{Kabirnezhad:2017jmf,Kabirnezhad:2020wtp,Kabirnezhad:2022znc}.
Therefore, when the MK model is implemented and validated in the future, $\thetacom$ can be used to study these resonance correlations.
This is particularly important for DUNE, where resonance production plays a major role.

One limitation of using $\thetacom$ is that the FSI effects causing variations in the measured hadronic kinematic distributions are entangled with resonance effects.
Hence, for the most sensitive study of resonance correlations, $\nu$-H selections are most suitable, as shown in Fig.~\ref{fig:enu-comp-h}.

Furthermore, $\thetacom$ measurements based on $\nu$-H selections can be compared directly between different fluxes, as shown in Fig.~\ref{fig:flux-comp}.
If high-purity $\nu$-H selections are achieved, $\thetacom$ measurements can provide a valuable pivot for cross-experiment comparison.

The importance of a high-purity $\nu$-H sample cannot be overstated.
For example, such events provide a unique opportunity for precise measurements of the axial current—a property unique to neutrinos.
These advantages have driven significant efforts to develop techniques for selecting a high-purity $\nu$-H sample, as demonstrated in Refs.~\cite{Lu:2015hea,MINERvA:2023avz,Baudis:2023tma}.
While Refs.~\cite{Baudis:2023tma} and \cite{MINERvA:2023avz} focus on neutrino and antineutrino pion-less events, respectively, Ref.~\cite{Lu:2015hea} pertains to the same event topology as that used for the COM total energy.
With new or future detectors—e.g., the SFGD and the 3DST of DUNE~\cite{DUNE:2021tad}—there is greater potential to implement novel techniques for selecting a high-purity $\nu$-H sample.
To this end, the sensitivity of the COM frame to FSI can be leveraged not only to study FSI but also to select events with minimal FSI.
Preliminary studies have shown that at the SFGD, a considerable number of carbon background events remain after optimal cuts on TKI variables—namely, $\dptt$ and $\dpt$.
Adding a cut on $\ecom$ can remove a significant portion of these background events, thereby achieving higher purity.
A more thorough investigation of this application of $\ecom$ is required to obtain a reliable assessment of the improvement in purity and will be explored in future studies.

\section{conclusion}
This study introduces a novel set of variables—namely, the COM angle and total energy.
These variables offer several advantageous properties, thereby motivating their use in future cross‐section measurements.
Such measurements have the potential to significantly improve our understanding of, and modeling of, FSI processes in neutrino–nucleus interactions.
The use of $\ecom$ to select a high-purity $\nu$-H sample will be explored in future work.
hen such a sample becomes available, the COM angle can shed light on resonance correlations and serve as a crucial pivot for cross‐experiment comparisons.

\section{Acknowledgement}
The author would like to thank Stephen Dolan for his constructive advice and helpful discussion.

% The \nocite command causes all entries in a bibliography to be printed out
% whether or not they are actually referenced in the text. This is appropriate
% for the sample file to show the different styles of references, but authors
% most likely will not want to use it.
% \nocite{*}

\bibliography{apssamp}% Produces the bibliography via BibTeX.

\end{document}